\documentclass[fleqn,usenatbib]{mnras}

\usepackage{newtxtext,newtxmath}
\usepackage{float}
\usepackage{placeins}
\usepackage{multirow}

\usepackage[T1]{fontenc}

\usepackage{babel}

\DeclareRobustCommand{\VAN}[3]{#2}
\let\VANthebibliography\thebibliography
\def\thebibliography{\DeclareRobustCommand{\VAN}[3]{##3}\VANthebibliography}

\usepackage{graphicx}	
\usepackage{amsmath,bm}	
\usepackage{hyperref}
\usepackage{physics}
\usepackage{mathtools}
\usepackage{tcolorbox}
\usepackage{enumitem}
\usepackage{mathrsfs}
\usepackage[sharp]{easylist}

\usepackage{xcolor}

\newcommand{\msun}{\mathrm{M}_{\odot}}
\newcommand{\rh}{r_{\mathrm{H}}}
\newcommand{\rhs}{r_{\mathrm{H,s}}}
\newcommand{\Eh}{E_{\mathrm{H}}}
\newcommand{\Ebin}{E_{\mathrm{bin}}}

\newcommand{\rp}{r_{\mathrm{p}}}

\newcommand{\mh}{m_{\mathrm{H}}}
\newcommand{\mbh}{m_\mathrm{BH}}
\newcommand{\mbin}{M_\mathrm{bin}}
\newcommand{\msmbh}{M_\mathrm{SMBH}}

\newcommand{\thefontsize}{The current font size is: \f@size pt}

\newlist{steps}{enumerate}{1}
\setlist[steps]{
  label=Step \Roman*,
  leftmargin=*,
  align=left
}

\tcbuselibrary{skins}

\newcommand{\ol}[1]{\begin{enumerate}#1\end{enumerate}}

\newcommand{\li}[1]{\item{#1}}

\newlength{\shiftwidth}
\defcitealias{Rowan_2023}{CR1}
\defcitealias{Rowan_2024}{CR2}
\defcitealias{Whitehead_2024I}{HW1}
\defcitealias{Whitehead_2024II}{HW2}

\begin{document}


\title[Adiabatic Simulations of BBH Formation in 3D]{3D Adiabatic Simulations of Binary Black Hole Formation in AGN Discs}

\author[H. Whitehead et al.]{
Henry Whitehead$^{1}$\thanks{E-mail: henry.whitehead@physics.ox.ac.uk}, Connar Rowan$^{2,3}$,
Bence Kocsis$^{2,4}$
\\
$^{1}$Department of Physics, Astrophysics, University of Oxford, Denys Wilkinson Building, Keble Road, Oxford OX1 3RH, UK\\
$^{2}$Rudolf Peierls Centre for Theoretical Physics, Clarendon Laboratory, University of Oxford, Parks Road, Oxford, OX1 3PU, UK \\
$^{3}$Niels Bohr International Academy, The Niels Bohr Institute, Blegdamsvej 17, DK-2100, Copenhagen , Denmark \\
$^{4}$St Hugh's College, St Margaret's Rd, Oxford, OX2 6LE, UK \\
}

\date{\today}

\pubyear{2023}

\label{firstpage}
\pagerange{\pageref{firstpage}--\pageref{lastpage}}
\maketitle

\begin{abstract}
We investigate close encounters between initially unbound black holes (BHs) in the gaseous discs of active galactic nuclei (AGN), performing the first 3D non-isothermal hydrodynamical simulations of gas-assisted binary BH formation. We discuss a suite of 135 simulations, considering 9 AGN disc environments and 15 BH impact parameters. We find that the gas distribution within the Hill sphere about an isolated embedded BH is akin to a spherically symmetric star with a low-mass convective envelope and a BH core, with large convective currents driving strong outflows away from the midplane. We find that Coriolis force acting on the outflow results in winds, analogous to cyclones, that counter-rotate with respect to the midplane flow within the Hill sphere. We confirm the existence of strong thermal blasts due to minidisc collisions during BH close encounters, as predicted in our previous 2D studies. We document binary formation across a wide range of environments, finding formation likelihood is increased when the gas mass in the Hill sphere is large, allowing for easier binary formation in the outer AGN disc. We provide a comprehensive overview of the SMBH's role in binary formation, investigating how binary formation in intermediate density environments is biased towards certain binary orientations. We offer two models for predicting dissipation by gas during close encounters, as a function of the ambient Hill mass alone, or with the periapsis depth. We use these models to motivate a prescription for binary formation likelihood that can be readily applied to Monte-Carlo simulations of AGN evolution.
\end{abstract}




\section{Introduction}

Active galactic nuclei (AGN) are promising locations for binary black hole (BBH) formation and mergers and may make significant contributions to the gravitational wave (GW) events detected by LIGO/VIRGO/KAGRA (LVK) \citep{McKernan+2012, McKernan+2014, Tagawa_2020, McKernan_2024}. BBH interactions in AGN are expected to be frequent due to the high density of compact objects in the galactic core which may then be captured into the AGN disc \citep{Bahcall_1976, Miralda_2000, Kennedy_2016, Bartos_2017, Panamarev_2018, Fabj_2020, Nasim_2023}. BHs may also form within the AGN itself due to gravitational instability in the outer disc \citep{Stone_2017, Secunda+2019}. The interaction frequency between BHs in AGN may be enhanced if the BHs are captured into migration traps, regions where there is a sign change in the migration driven by gas torques \citep{Paardekooper_2006, Bellovary_2016, Secunda_2019, Secunda_2020, Grishin_2024}. During close encounters, BHs can form stable binaries by a variety of means. Gravitational waves alone can form a binary over a very small capture cross-section \citep{OLeary+2009}; this cross-section can be enhanced by Jacobi encounters, where the SMBH drives the binary through multiple encounters before ionisation \citep{Boekholt_2022}. If another BH is present, three-body encounters can also drive binary formation and hardening \citep{Leigh_2018, Samsing_2022}. This study focuses on a third case, where gas gravity dissipation \citep{Tagawa_2020, Li_2023, Rowan_2023, Rowan_2024, Whitehead_2024I, Whitehead_2024II} drives binary formation. 

Along with the potential to contribute to the GW rate, the AGN channel is of specific interest due to the potential for associated electromagnetic emission at merger \citep{McKernan_2019, Graham_2020, Kimura_2021, Wang_2021, Chen_2023, Tagawa_2023_merge}, or even at the moment of binary formation \citep{Whitehead_2024II}. 

Accurate modelling of BBH formation by gas gravitation in AGN is crucial to informing studies attempting to calculate the contributions of the AGN channel to the overall LVK rates as a function of source parameters. Previous studies have implemented semi-analytical methods \citep{Tagawa_2020} or Monte-Carlo simulations \citep{McKernan_2024_I, Delfavero_2024}, evolving the population of BH and BBH binaries to estimate both the frequency of mergers and the properties of the merging binaries. Mergers in these simulations may occur for pre-existing binaries, but generally are dominated by those binaries formed within the disc by gas-capture \citep{Tagawa_2020}. While older models considered gas dynamical friction timescales to calculate the likelihood of binary formation, more recent studies have directly implemented models informed by hydrodynamical simulations \citep{Rowan_2025a}. 

Simulating BBH formation in AGN warrants care for a variety of reasons. Notably, studies must consider a large range of length scales in order to resolve both the circum-BH flow and the motion of the BHs through the disc. There are also many physical processes that can affect the gas dynamics including viscosity, equation-of-state, radiation and magnetism. Additionally, there are a wide variety of initial conditions pertinent to the encounter outcome such as the ambient disc conditions and BH trajectories. Previous works have approached the problem with a variety of numerical methods, including semi-analytical gas dynamical friction \citep{delaurentiis_2022, Qian_2023, Rozner+2023}, 3D isothermal smoothed particle hydrodynamics \citep{Rowan_2023, Rowan_2024} and 2D Eulerian hydrodynamic codes \citep{Li_2023, Whitehead_2024I, Whitehead_2024II}. Of significant importance to the AGN channel are studies of pre-existing binaries that study the binary evolution over many orbits \citep{Baruteau_2011, LiLai_evo, LiLai_eos, LiLai_visc, Dittmann_2024, Calcino_2024}. \citet{Mishra_2024} performed simulations of single and binary black holes in an isothermal turbulent magnetised AGN disc, making it the first study in this field to include magnetic fields. Recently, \citet{Rowan_2025b} and \citet{Wang_2025} extended the analysis of embedded binaries to consider binary-single scattering events under a hydrodynamical treatment.

This paper features the most realistic treatment of the embedded formation scenario to date, modelling binary interactions between two initially isolated BHs embedded within four different AGN disc environments to form a suite of 135 3D viscous adiabatic simulations. We examine the behaviour of the circum-BH flow when the BHs are distant, documenting the gas morphologies. We track the binary components through close-encounter and report where binary formation occurs, providing a large number of simulations spanning a variety of disc environments and encounter types. In analysing the statistics of our large suite of simulations, we provide a hydrodynamically-informed model for binary black hole formation in generalised AGN environments. We first describe the computational methodology in Section~\ref{sec:comp_method}, followed by the initial conditions for the simulations in Section~\ref{sec:ic}. In Section~\ref{sec:fid_results} we present the results of a fiducial simulation, describing the pre-encounter gas morphology and the chronology of a binary formation event. In Section~\ref{sec:param} we widen our scope to the full simulation suite, comparing across AGN environment and a variety of BH trajectories. We discuss the findings of the study in Section~\ref{sec:discuss} and summarise our conclusions in Section~\ref{sec:conclusions}.

\section{Computational Methods}
\label{sec:comp_method}
This work builds directly onto our two previous works \citep{Whitehead_2024I, Whitehead_2024II}, hereafter \citetalias{Whitehead_2024I} \& \citetalias{Whitehead_2024II} respectively, which simulated BBH formation in 2D: here we extend the domain to 3D but maintain much of the core computational setup. We use the Eulerian general relativistic magneto-hydrodynamics (GRMHD) code \texttt{Athena++} \citep{Stone_2020} to perform our hydrodynamical simulations, neglecting any effects associated with gas self-gravity, relativity, magnetism and radiative transfer. We utilise a second-order accurate van Leer predictor-corrector integrator with a piecewise linear method (PLM) spatial reconstruction and the Harten-Lax-van Leer-Contact (HLLC) Riemann solver. We simulate a 3D cuboid patch of disc (the ``shearing box'') in a corotating frame about the SMBH. The standard length scales in the shearing frame are the single and binary Hill radii: $\rhs$ and $\rh$ respectively,
\begin{equation}
    \rhs = R_0\left(\frac{m_\mathrm{BH}}{3\msmbh}\right)^\frac{1}{3}, \quad \quad \rh = R_0\left(\frac{\mbin}{3\msmbh}\right)^\frac{1}{3}.
\end{equation}
Here $R_0$ is the distance from the SMBH with mass $\msmbh$, with $\mbh$ and $\mbin$ the single and binary BH masses respectively. Throughout the paper, subscripts $_\mathrm{H,s}$ refer to quantities appropriate for isolated BHs and subscripts $_\mathrm{H}$ refer to binary quantities.

\subsection{The Shearing Box}
\label{sec:shear_box}
The shearing box is a non-inertial domain that co-rotates with the global AGN disc at a radius $R_0$. In this local region, we introduce Cartesian coordinates $\{x,y,z\}$ related to the global position $\bm{r}$ by
\begin{equation}
    \bm{r} = 
    \begin{pmatrix}
        R \\
        \phi \\
        z
    \end{pmatrix}
    =
    \begin{pmatrix}
        R_0 + x \\
        \Omega_0 t + \frac{y}{R_0} \\
        z
    \end{pmatrix},
\end{equation}
where $\Omega_0 = \sqrt{\frac{G\msmbh}{R_0^3}}$ is the angular frequency of the frame about the SMBH. In the shearing box, all bodies experience additional accelerations due to fictitious Coriolis, centrifugal forces and the vertical component of the SMBH gravity, we define the sum of these as
\begin{equation}
    \label{eq:a_SMBH}
    \bm{a}_\text{SMBH} = 2\bm{u} \times \Omega_0 \hat{\bm{z}} + 2q\Omega_0^2 \bm{x} - \Omega_0^2 \bm{z},    
\end{equation}
where we have introduced the shear rate $q = -\frac{\text{d}\ln \Omega}{\text{d} \ln R} = \frac{3}{2}$ and $\bm{u}$, the body's velocity in the rotating frame. These fictitious forces vanish for equilibrium trajectories satisfying
\begin{equation}
    \bm{u}_\text{eq} = 
    \begin{pmatrix}
        u_x \\
        u_y \\
        u_z
    \end{pmatrix}
    =
    \begin{pmatrix}
        0 \\
        - q\Omega_0 x \\
        0
    \end{pmatrix} ,\quad z = 0,
\end{equation}
where here $\bm{u}$ is expressed in Cartesian form. These equilibrium trajectories are straight lines in the shearing frame: in the global frame they are zero-inclination, circular orbits of varying radii.

\subsection{Gas Dynamics}
\label{sec:gas_dyn}
Gas evolves within the shearing box according to the Navier-Stokes equations, with additional accelerations associated with the SMBH (see Section~\ref{sec:shear_box}).
\begin{equation}
    \centering
    \frac{\partial \rho}{\partial t} + \nabla \cdot \left(\rho \bm{u}\right) = 0,
\end{equation}
\begin{equation}
    \label{eq:mom_evo}
    \frac{\partial \left(\rho \bm{u}\right)}{\partial t} + \nabla \cdot \left(\rho \bm{u} \bm{u} + P \bm{I} + \bm{\Pi}\right) = \rho\left(\bm{a}_{\text{SMBH}} + \bm{a}_\text{BH}\right),
\end{equation}
where we have introduced $\rho$, $P$, $\bm{u}$ and $\bm{\Pi}$ as the gas density, pressure, velocity and viscous stress tensor with components
\begin{equation}
    \Pi_{ij} = \rho \nu \left(\frac{\partial u_i}{\partial x_j} + \frac{\partial u_j}{\partial x_i} - \frac{2}{3}\delta_{ij}\nabla \cdot \bm{u}\right),
\end{equation}
for a given kinematic viscosity $\nu$. For simplicity, we adopt a single static and homogeneous value for $\nu$ as given by the ambient kinematic viscosity in the local AGN disc, such that $\nu = \alpha \Omega_0 H_0^2$ where $H_0$ is the ambient disc scale height and $\alpha$ is the Shakura-Sunyaev viscosity coefficient. We have also added a term $\vb{a}_\mathrm{BH}$ to Equation~\ref{eq:mom_evo} which accounts for gravitation by any stellar mass BHs within the simulation: 
\begin{equation}
    \boldsymbol{a}_\text{BH} = -\nabla \phi_\text{BH}(\boldsymbol{r}) = \sum^{n_\text{BH}}_{n=1} m_{\text{BH},n} \,g\left(\frac{\boldsymbol{r}-\boldsymbol{r}_n}{h}\right)
\end{equation}
where $m_{\mathrm{BH},n}$ and $\bm{r}_n$ are the mass and position of the $n^\mathrm{th}$ BH and $h = 0.025\rhs$ is the smoothing length for the gravitational spline kernel, $g(\bm{\delta}$) \citep[][Appendix A]{Price_2007}.
\begin{equation}
    \label{eq:kernel}
    g(\boldsymbol{\delta}) = -\frac{G}{h^2}\hat{\boldsymbol{\delta}}
    \begin{cases}
    \frac{32}{3}\delta - \frac{192}{5}\delta^3 + 32\delta^4 & 0 < \delta \le \frac{1}{2} \\
    -\frac{1}{15\delta^2} + \frac{64}{3}\delta - 48\delta^2 + \frac{192}{5}\delta^3 - \frac{32}{3}\delta^4 & \frac{1}{2} < \delta \le 1 \\
    \frac{1}{\delta^2} & \delta > 1
    \end{cases}
\end{equation}
This kernel reproduces the gravitational force exactly for $r \geq h$, and smoothly transitions to zero at $r=0$. The BHs are introduced steadily into the simulation, such that their mass grows from zero over a period of $t_\mathrm{grow} = 0.25P_\mathrm{SMBH}$ where $P_\mathrm{SMBH} = 2\pi\Omega_0^{-1}$; this is done to give the gas time to adjust to the BHs' presence and reduce aggressive shock heating at early times. Mass growth and softening are applied only to BH-gas interactions; the BH-BH gravitation is unsoftened and uses the true masses.

Unlike this paper's predecessor \citetalias{Whitehead_2024II} which considered a mixture of radiation and gas in 2D, here we focus only on adiabatically evolving gas in 3D. Removing the radiation component reduces computational expense and proves to be numerically stabler. However, in doing so, we have removed a pressure term from our fluid evolution. Furthermore, in 3D we are unable to implement the on-the-fly cooling terms used in 2D. Such approximations are inaccessible to our 3D simulations, where accurate cooling dynamics would require a full radiative transport treatment which we neglect here due to computational cost. We reason that given the generally high optical depth of these systems near the midplane, neglecting radiative transport and therefore cooling from this study allows for a reasonable first approximation of the hydrodynamics close to the BHs. We leave more complete computational representations of these flows for future studies. As such the energy of the gas evolves as
\begin{equation}
    \label{eq:energy_cons}
    \frac{\partial E}{\partial t} + \nabla \cdot \left[\left(E+P\right)\bm{u} + \bm{\Pi} \cdot \bm{u}\right] = \rho \bm{u} \cdot \left(\bm{a}_\text{SMBH} + \bm{a}_\text{BH}\right).
\end{equation}
Here $E$ represents the total fluid energy per unit volume, separable into internal and kinetic components
\begin{equation}
    E = U + K = U + \frac{1}{2}\rho \bm{u} \cdot \bm{u}.
\end{equation}
We assume the gas behaves ideally, such that the gas pressure $P$ and internal energy $P$ can be expressed as 
\begin{align}
    P &= \frac{k_\text{B}}{\mu_p m_u}\rho T, \\
    U &= \frac{P}{\gamma -1} = \frac{3}{2}\frac{k_\text{B}}{\mu_p m_u}\rho T,
\end{align}
where $k_B$, $\mu_p$, $m_u$, $\gamma = \frac{5}{3}$ are the Boltzmann constant, average molecular weight, atomic mass constant and the adiabatic constant for monatomic gas respectively. We model the AGN gas as a fully ionised mixture of H, He with mass fractions $(X,Y) = (0.7,0.3)$ such that the average molecular weight is $\mu_p = \frac{8}{13}$. 

\subsection{Adaptive Mesh Refinement}
\label{sec:amr}
In order to significantly reduce computational expense while still maintaining a high spatial resolution in the regions of interest we implement an adaptive mesh refinement (AMR) routine that increases the number of cells near the BH. Mesh regions nearer to the BHs are more refined, with each extra refinement level halving the cell side length (increasing the number of cells in said region eight-fold). The AMR criterion is preferentially biased towards refinement in the disc plane (as opposed to in the $z$ direction) by a factor $z_\text{bias} = 2$, this is implemented via an effective distance $r_\text{eff}^2 = (x-x_\text{BH})^2 + (y - y_\text{BH})^2 + z_\text{bias}^2 (z - z_\text{BH})^2$. If $r_\text{eff} < 0.2\rhs$, the mesh is maximally refined to 6 levels above the root grid. If $0.2\rhs < r_\text{eff} < 0.5\rhs$, the mesh is refined to 5 levels above root. Outside this region, the refinement transitions smoothly to the base level. As the BHs propagate through the simulation, these hyper-refined regions will move with them. This scheme allows us to focus the majority of the computation power on resolving the BH minidiscs, while still simulating the large scale flow structures at a lower resolution. 

\section{Initial Conditions}
\label{sec:ic}
\begin{figure}
    \includegraphics[width=\columnwidth]{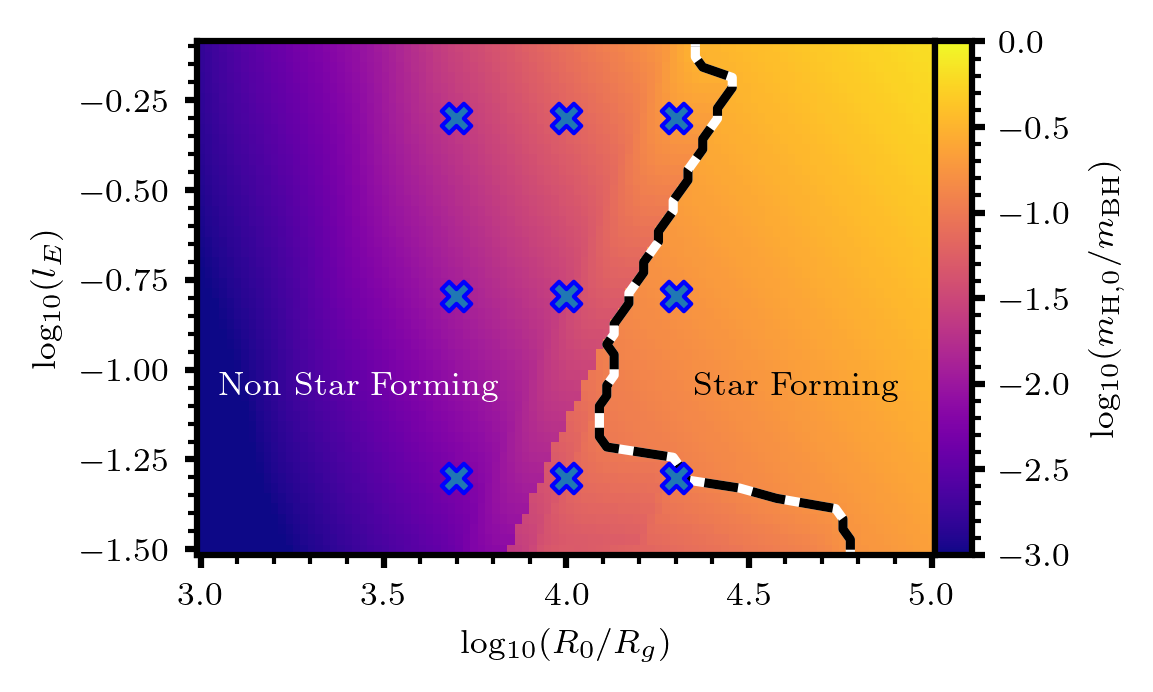}
    \caption{Ambient Hill masses $m_{\mathrm{H},0}$ (Eq.~\ref{eq:mh0}) for a range of AGN environments: varying position in the disc $R_0$, and Eddington fraction $l_E$. All other AGN properties are held fixed as in Table~\ref{tab:static}. Superimposed as crosses are the 9 AGN environments considered in this paper. We expect binary formation to be biased towards regions of higher ambient Hill mass, hence our focus on regions in the outer AGN disc where the Hill radius is larger.}
    \label{fig:amb_mass}
\end{figure}

In this paper, we consider the dynamics of binary black hole formation between two $25M_\odot$ BHs in 9 different AGN disc environments, with the initial hydrodynamic states computed using the pAGN pipeline \citep{Daria_2024} which implements a vertically one-zone axisymmetric equilibrium Shakura-Sunyaev $\alpha$ disk \citep{Shakura_1973} assuming self-regulating equilibrium with an extra source of heat (and pressure) in the star-forming regions \citep{Sirko_Goodman2003,Thompson+2005}. We consider 3 logarithmically spaced values for both the radial location within the AGN disc $R_0 \in [5\times 10^3, 10^4, 2\times 10^4]R_g$ and the total AGN disc Eddington fraction $l_E \in [0.05, 0.16, 0.5]$. Table~\ref{tab:static} records the AGN parameters fixed across all simulations, Table~\ref{tab:amb} records the ambient gas properties for each of the 9 AGN environments. In this study we aim to explore a range of regions with different binary formation rates. \citetalias{Whitehead_2024I} found the minidisc mass to be the key determinant for successful binary formation, as more massive minidiscs resulted in greater gas dissipation during close encounters. The minidisc mass is positively correlated with the ambient Hill mass:
\begin{equation}
    \label{eq:mh0}
    m_\mathrm{H,0} = \iiint_{r<\rh} \rho dV \sim \Sigma_0 \pi \rh^2 = 2\pi \rho_0 H_0 \rh^2,
\end{equation}
where $\Sigma_0$, and $H_0$ are the vertically integrated ambient density and disc scale height respectively. Throughout the parameter space explored in this study, $\rh > H_0$. The 9 environments selected represent a wide range of ambient Hill masses, with $m_{\mathrm{H},0} \in \left[0.005, 0.2\right]$. These environments are located close to the boundary of the star forming region in the outer AGN disc, see Figure~\ref{fig:amb_mass}. Binary formation may still be viable deeper within the AGN disc, but due to the linear dependence of $\rh$ on $R_0$ the volume that the BH can trap mass within decreases severely, resulting in diminishing $m_{\mathrm{H},0}$. We note that one of our simulation environments $(R_0, l_E) = (2\times 10^4R_g, 0.16)$ lies within the star-forming region of the outer AGN disc, where the initial conditions for the gas are less well understood. This simulation receives the same modelling treatment as the other models regardless.
Within each of the 9 AGN environments, we run 15 simulations of BH-BH close encounters, varying the impact parameter $b$ linearly between 1.7 - 2.4$\rh$. Binary formation may be viable outside of this range, especially when $m_{\mathrm{H},0}$ is high, but we implement a relatively restricted range to allow for better resolution in $b$. The BH trajectories at close encounter are very sensitive to the initial impact parameter, as the presence of the SMBH makes all encounters 3-body interactions.

\begin{table}
    \centering
    \begin{tabular}{| c c c c c|}
    \hline
    $\msmbh$ & $\alpha$ & $\epsilon$ & $X$ & $b$ \\ [0.5ex]
    \hline
    $4 \times 10^6 M_\odot$ & 0.1 & 0.1 & 0.7 & 0 \\ [1ex]
    \hline
    \end{tabular}
    \caption{AGN disc parameters held constant for all models discussed in this paper. From left to right, the quantities are SMBH mass, Shakura-Sunyaev viscosity coefficient, radiative efficiency, hydrogen abundance and switch of viscosity-pressure relation $\nu = \alpha \Omega^{-1} P_g^b P^{1-b}$. These quantities are fed to the pAGN pipeline \citep{Daria_2024} when generating ambient disc states, which then set the hydrodynamic initial conditions of each simulation (see Table~\ref{tab:amb}).}
    \label{tab:static}
\end{table}

\begin{table*}
    \centering
    \begin{tabular}{| c c c c c c c|}
    \hline
    $R_0$ $[R_g]$ & $l_E$ & $\rho$ $[\mathrm{gcm}^{-3}]$ & $\Sigma$ $[\mathrm{gcm}^{-2}]$ & $T$ $[\mathrm{K}]$  & $c_s$ $[\mathrm{cms}^{-1}]$ & $\frac{H}{R}$ \\ [0.5ex]
    \hline
    $5\times 10^3$ & 0.05 & $1.81 \times 10^{-10}$ & $9.55 \times 10^3$ & $1.15 \times 10^4$ & $1.34 \times 10^6$ & $4.48 \times 10^{-3}$ \\
    $5\times 10^3$ & 0.16 & $3.22 \times 10^{-10}$ & $2.07 \times 10^4$ & $1.60 \times 10^4$ & $1.63 \times 10^6$ & $5.44 \times 10^{-3}$ \\ 
    $5\times 10^3$ & 0.50 & $4.92 \times 10^{-10}$ & $4.02 \times 10^4$ & $2.23 \times 10^4$ & $2.08 \times 10^6$ & $6.91 \times 10^{-3}$ \\ 
    $1\times 10^4$ & 0.05 & $4.67 \times 10^{-10}$ & $2.55 \times 10^4$ & $1.78 \times 10^3$ & $4.90 \times 10^5$ & $2.31 \times 10^{-3}$ \\
    $1\times 10^4$ & 0.16 & $9,96 \times 10^{-11}$ & $1.34 \times 10^4$ & $9.37 \times 10^3$ & $1.21 \times 10^6$ & $5.70 \times 10^{-3}$ \\ 
    $1\times 10^4$ & 0.50 & $1.57 \times 10^{-10}$ & $2.66 \times 10^4$ & $1.33 \times 10^4$ & $1.52 \times 10^6$ & $7.16 \times 10^{-3}$ \\ 
    $2\times 10^4$ & 0.05 & $9.58 \times 10^{-11}$ & $1.26 \times 10^4$ & $1.28 \times 10^3$ & $4.16 \times 10^5$ & $2.78 \times 10^{-3}$ \\
    $2\times 10^4$ & 0.16 & $9.58 \times 10^{-11}$ & $1.85 \times 10^4$ & $2.78 \times 10^3$ & $6.13 \times 10^5$ & $4.09 \times 10^{-3}$ \\ 
    $2\times 10^4$ & 0.50 & $6.90 \times 10^{-11}$ & $2.18 \times 10^4$ & $8.82 \times 10^3$ & $1.00 \times 10^6$ & $6.67 \times 10^{-3}$ \\ [1ex]
    \hline
    \end{tabular}
    \caption{Gas properties for all 9 AGN environments discussed in this paper. From left to right the quantities are the distance from the SMBH, the SMBH Eddington fraction, midplane density, surface density, midplane temperature, sound speed and disc aspect ratio. These states are generated by feeding the values from Table~\ref{tab:static} to the pAGN pipeline \citep{Daria_2024}.}
     \label{tab:amb}
\end{table*}

\subsection{Simulation Dimensions}
\label{sec:dimensions}
When simulating potential binary formation events, the BHs are initialised on circular orbits about the central SMBH separated by an impact parameter $b$. The centre-of-mass for the BHs is positioned at the centre of the shearing frame. Each of our interactions are simulated in a box with dimensions $x \in \left[-4\rh, 4\rh\right]$, with a $y$ extent sufficient to allow for a pre-encounter flight time equal to $P_\text{SMBH} = 2\pi\Omega_0^{-1}$, the period of frame about the SMBH . The box has a vertical extent of $z \in \left[-10H_0, 10H_0\right]$; such a large vertical extent ($z_\mathrm{max}\sim 0.05R_0$) is required to document the large outflows generated, but the additional cost associated with this extension can be mitigated using adaptive mesh refinement (see Section~\ref{sec:resolution}).

\subsection{Vertical Structure}

For simplicity, we initialise our system as homogeneous in the $x-y$ plane (i.e. neglecting the small change in temperature with changing radius), but with vertical stratification. Vertical hydrostatic equilibrium requires a pressure gradient to balance against the SMBH gravity, such that
\begin{equation}
    \partial_z P + \rho \Omega_0^2 z = 0.
\end{equation}
We adopt an isothermal atmosphere, such that the initial density profile is given by the standard Gaussian,
\begin{equation}\label{eq:rho}
    \rho(z) = \rho_0 \exp(\frac{-z^2}{2H_0^2}),
\end{equation}
where $H_0 = c_s/\Omega_0$ and $c_s^2\equiv P/\rho$. As the existence of large-scale outflows motivates the use of a box with significant vertical extent, we are forced to include regions where the AGN disc density would be vanishingly small. This is numerically problematic and density values can approach machine precision, to combat this we impose a density floor in the initial conditions of all simulations, such that $\rho \geq \rho_{\rm b} = 10^{-7}\rho_0$. This leads to a violation of hydrostatic equilibrium at large $z$, but as this takes place in regions of negligible density, we do not find it to have significant effect on the large scale hydrodynamics. 

\subsection{Resolution}
\label{sec:resolution}
The base resolution of each simulation is set such that there are 8 cells per Hill radius, with the base cell size $\Delta x_{n=0} = \frac{\rh}{8}$. The domain is then scaled to accommodate this resolution, domain sizes range from $[n_x, n_y, n_z] = [64,192,48]$ to $[n_x, n_y, n_z] = [64,256,80]$, where $n_x$, $n_y$, $n_z$ are the number of cells along the $\hat{\bm{x}}$, $\hat{\bm{y}}$ and $\hat{\bm{z}}$ axes respectively. The domain size is dependent on the disc scale height and time to encounter. On top of the base, we apply 6 levels of mesh refinement (for 7 levels total), such that the minimum cell size close to BHs is $\Delta x_{n=6} = \frac{\rh}{512}$. Under this treatment, each simulation hosts $~3\times 10^7$ cells, with $\sim10^7$ cells within $\rhs$ of each BH. See Section~\ref{sec:amr} for a full description of the refinement routine. Such a fine resolution is required to resolve the complex pressure/gravitational gradients in the minidiscs. When compared to an exploratory simulation with 8 levels of mesh refinement and half the softening length ($h=0.0125\rhs$), we observe identical flow morphology. It is likely that increasing the resolution further would more accurately reproduce the fluid behaviour deep within the minidiscs, but we expect the resolution used here to be sufficient for binary formation studies. Studies focusing on the evolution of binary elements post formation may wish to implement a finer resolution; \citet{Rowan_2025b} found that higher resolutions were required to accurately model binary-single scatterings in 2D.

\section{Fiducial Results}
\label{sec:fid_results}
Before considering the full suite of simulations we focus on a specific case study featuring successful binary formation, the system with $\left(l_E, R_0, b\right) = \left(0.16, 10^4 R_g, 2.30 \rh\right)$. We begin with a discussion of the circum-BH structure in isolation, before moving to the close-encounter between BHs and resultant binary formation. In Section~\ref{sec:param} we expand our analysis to the rest of the simulation suite and compare across the parameter space.

\subsection{Gas Morphology in Isolation}
\label{sec:isolation}

As the simulation evolves, the BHs draw gas from the ambient AGN disc into their Hill sphere, constructing a minidisc. In \citetalias{Whitehead_2024II}, we found the mass of the minidisc to be dependent on the equation of state; isothermal systems were able to grow $\sim10-15\%$ more massive minidiscs due to the reduced fluid pressure. In 3D we observe thermal winds emanating vertically away from the disc midplane. These winds result in mass being stripped from the Hill sphere and ejected to heights significantly greater than the ambient AGN disc height $H_0$. In discussion of the circum-BH flow, we adopt cylindrical coordinates $\{R, \phi, z\}$ centred on each BH, with velocities $\bm{u}$ in the frame of the BH.

\subsubsection{Wind Growth and Mass Evolution}

\begin{figure*}
    \includegraphics[width=2\columnwidth]{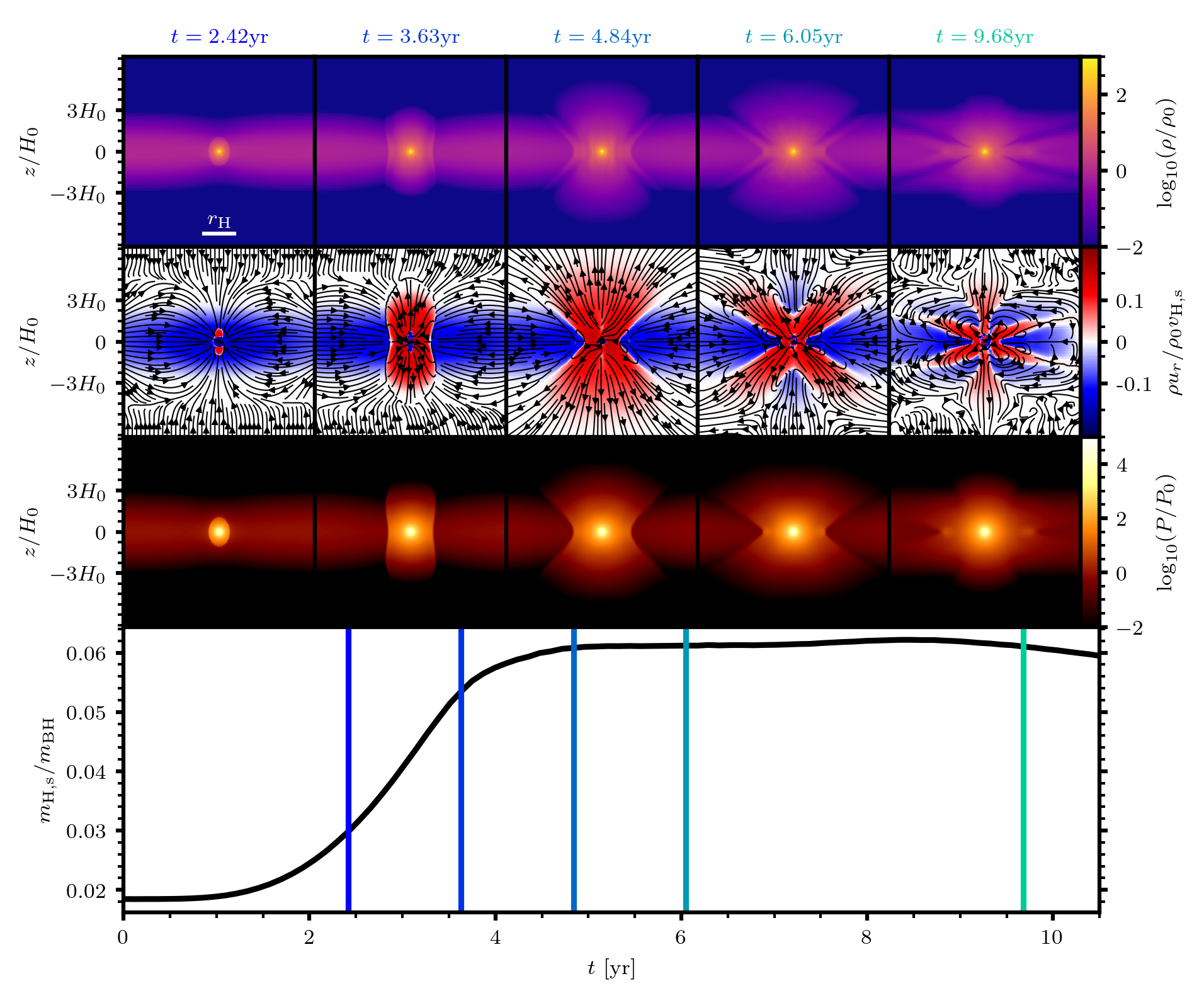}
    \caption{Growth of a minidisc wind in the fiducial simulation depicting slices in the $y$-$z$ plane off the logarithmic gas density, the radial mass flux, gas pressure. In the lower panel, we document the evolution of the Hill mass. The wind initially develops as a hot bubble which breaks out vertically from the disc and then expands laterally to form a wide conical opening in the disc. By $t=5$yr, the inflow and outflow of gas into the Hill sphere has stabilised, resulting in an approximately constant Hill mass around $m_{\mathrm{H},s} \simeq 0.06\mbh$. As the strength of the wind dies, the conical wind collapses into an X-shaped outflow with inflow directly above and below the BH. At late times, this collapse can rebound, resulting in a new outflow component directly above/below the BH.}
    \label{fig:wind_growth}
\end{figure*}

We describe the formation of a minidisc wind while the BHs are in isolation, with aid of Figure~\ref{fig:wind_growth}. Once the simulation starts, gas gravitates towards the BH where it is shock heated, resulting in a hot bubble of gas in the Hill sphere. Thermal pressure drives this gas outwards, with the preferential direction vertically upwards due to the fall-off of gas density and pressure with height. This thermal wind breaks out from the AGN disc and is cast high into the disc atmosphere, with significant density ($\rho \sim 0.2\rho_0$) reaching $z > 3H_0$. Post-breakout, the wind widens, developing a pseudo-conical region directly above/below the minidisc. We observe a fall-back flow from large $z$ following the initial breakout, compressing the conical wind into an X-shaped outflow. Self-collision in the wind drives a vertical outflow component from directly above/below the BH at late times.

\begin{figure}
    \includegraphics[width=\columnwidth]{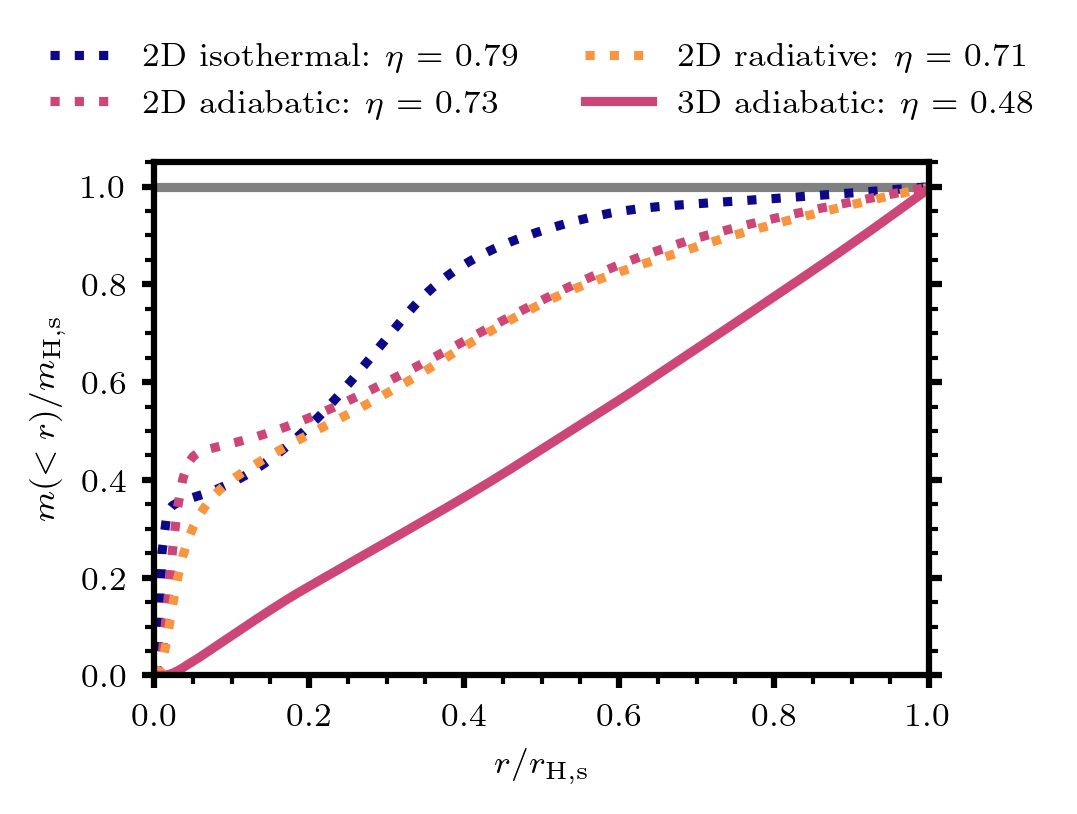}
    \caption{Cumulative minidisc mass distributions $m(<r)$ and compactnesses $\eta$, comparing between the 3D fiducial system and three 2D systems from \citetalias{Whitehead_2024II}. While all 2D systems exhibit comparably compactnesses of $\eta \sim 0.7-0.8$, the 3D system is substantially less compact. For the 2D systems, the half-mass-radius for the minidisc is around $0.2\rhs$, for the 3D system it is around $0.6\rhs$.}
    \label{fig:cmdf}
\end{figure}

The presence of a thermal wind has a significant effect on the minidisc, stripping mass and ejecting it from the Hill sphere. As shown in Figure~\ref{fig:wind_growth} this results in a plateau in the Hill mass when the mass inflow and outflow balances. In the fiducial case the Hill mass stabilises around $m_{\mathrm{H},s} = 0.06\mbh$. Compared to the 2D results of \citetalias{Whitehead_2024II}, our 3D minidiscs are about half as massive (when scaled by the ambient Hill mass). While in principle this could result in ineffective dissipation by gas gravity during binary interactions, the plateau mass is sufficient high for many of the systems studied here. It is not obvious how the inclusion of cooling or radiation effects should effect the height, or indeed existence, of this plateau mass. Cooling will likely act to increase this mass by decreasing the gas temperature, but including forces from radiation pressure may result in stronger winds driving more mass from the Hill sphere. See Section~\ref{sec:env_var} for more details on the relationship between the ambient Hill mass and the resultant Hill mass at close encounter. When considering the morphology of the minidiscs we introduce a parameter $\eta$ to describe the compactness of the gas within the Hill sphere.
\begin{equation}
    \eta = 1 - \frac{\int_{r<\rh} \rho r \mathrm{d}V}{\rh \int_{r<\rh} \rho \mathrm{d}V}
\end{equation}
The compactness $\eta$ ranges from 0 (minimal compactness, all gas in a shell at $r=\rh$), to 1 (maximal compactness, all gas in a point at $r=0$). In Figure~\ref{fig:cmdf} we apply this metric to the fiducial minidisc and to 2D minidiscs drawn from \citetalias{Whitehead_2024II}, plotting the cumulative mass distributions. In the 2D models, compactness is slightly reduced when moving from isothermal to adiabatic to radiative equations of state, as extra pressure support is provided. A more dramatic change is seen when moving to 3D, with $\eta \sim 0.5$ compared to $\eta \sim 0.7-0.8$ in 2D. The 3D system is substantially less compact than the 2D system, with a density profile closer to a spherically symmetric distribution than a disc (see Section~\ref{sec:bh_star}).

\subsubsection{Black Hole Stars in AGN}
\label{sec:bh_star}
\begin{figure*}
    \includegraphics[width=2\columnwidth]{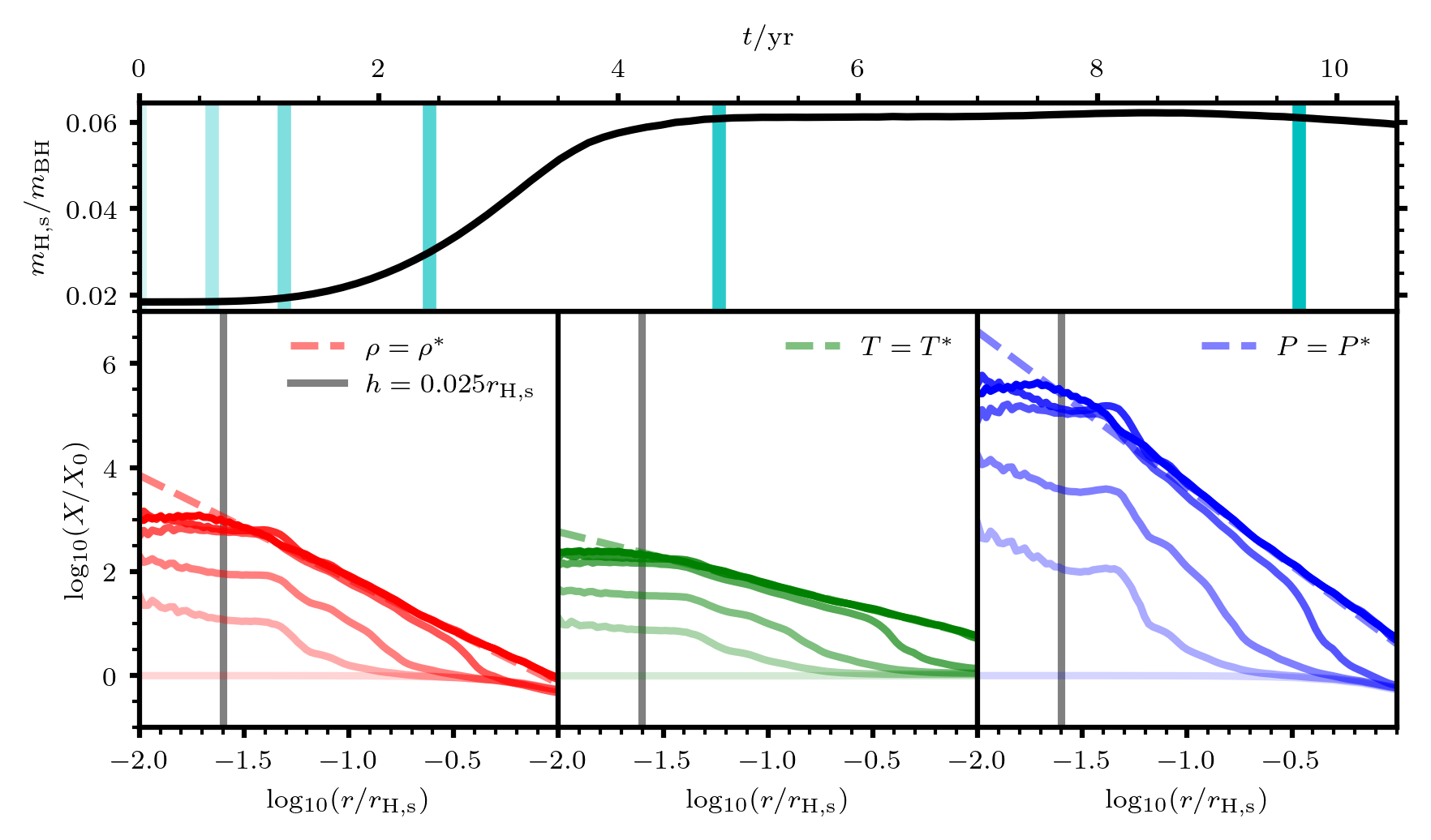}
    \caption{Radial distributions of gas properties within the Hill sphere for the fiducial system, averaging over all angles and volume weighting. In the top panel, the evolution of the Hill mass with time. In the lower panels, the time evolution of the gas density $\rho$, temperature $T$ and pressure $P$. Increasing line opacity indicates a later time, marked by the vertical lines in the top panel. By $t\sim5$yr, the system has reached the steady-state solution described by Equations~\ref{eq:ss_rho}-~\ref{eq:ss_P}.}
    \label{fig:radial}
\end{figure*}

Our fiducial simulation shows that the BHs quickly build a spherical gaseous envelope of mass $0.06\mbh$ within a radius $r_{\rm H,s}$. After 5 years, the compact $25\msun$ BH core is surrounded by a 10 AU radius gaseous sphere of mass $1.5\msun$. Figure~\ref{fig:radial} shows radial profiles for the gas properties within the Hill sphere, averaging over all angles and volume weighting each property. We find that outside of the smoothing radius $h=0.025\rhs$, the radial profiles evolve to steady power-law solutions by $t\sim 5$yr.
\begin{align}
    \label{eq:ss_rho}
    \rho^*(x) &= \rho_\mathrm{H}x^{-2}, \quad \rho_\mathrm{H} = \frac{m_\mathrm{H,s}}{4\pi \rhs^3}\,,\\.
    \label{eq:ss_T}
    T^*(x) &= T_\mathrm{H} x^{-1}, \quad T_\mathrm{H} = \frac{G\mbh \mu m_p}{3k \rhs}\,,\\
    \label{eq:ss_P}
    P^*(x) &= P_\mathrm{H}x^{-3}, \quad P_\mathrm{H} = \frac{G\mbh m_\mathrm{H,s}}{12\pi \rhs^4}\,.
\end{align}
Here $x = {r}/{\rhs}$ and $\rho_\mathrm{H}$, $T_\mathrm{H}$, $P_\mathrm{H}$ are the characteristic density, temperature and pressure for the Hill sphere, respectively. These forms can be derived analytically by considering spherically symmetric solutions to the equation for hydrostatic equilibrium about a point mass $\mbh$, neglecting gas self-gravity and frame forces (see Appendix~\ref{sec:virial}). This solution is unstable to convection; the Schwarzschild criterion for convective stability requires
\begin{equation}
    \partial_r K(r) < 0
\end{equation}
where $K(r)$ describes the entropy of the system \citep{Schwarzschild_2015}. However for the temperature and density profile represented in Equations~\ref{eq:ss_rho} and~\ref{eq:ss_T},
\begin{align}
    K(r) &\equiv P(r)\rho(r)^{-\gamma} \propto r^{2\gamma-3} \propto r^\frac{1}{3} \\
    &\implies \partial_r K(r) > 0
\end{align}
The Hill sphere's convective instability allows for an outflow structure despite satisfying hydrostatic equilibrium. As such, the gas within the Hill sphere behaves like a star composed of a massive $25M_\odot$ BH core and a low-mass $1.5M_\odot$ convective envelope. The study of stars within AGN discs is an active field of research \citep{Goodman_2004, Cantiello_2021, Jermyn_2021, Dittmann_2021, Jermyn_2022, Chen_2024a, Fabj_2024, Chen_2024b}. These AGN stars are commonly modelled as self-gravitating $n=3$ polytropes and can grow to $M_* = 100M_\odot$ \citep{Cantiello_2021}. In this study our stellar mass is dominated by the compact BH core, resulting in an envelope whose pressure-density profile mimics an $n=2$ polytrope, though caution is warranted when comparing to polytrope forms, as those equations assume that the gas is self-gravitating. Small scale BH physics such as accretion and feedback are neglected in our simulation, which have the potential to change the radial profiles. Given that we neglect accretion in the simulation the inflow and outflow rates in and out of the Hill sphere ultimately cancel. We note that the rate of accumulation of mass into the Hill sphere, 
$\dot{m}_\mathrm{H,s} \sim 0.2 M_\odot \mathrm{yr}^{-1} > 10^5\dot{M}_\mathrm{Edd}$ over the first 5 years of evolution in the simulation, is in significant excess of the Eddington mass accretion rate onto the BH $\dot{M}_\mathrm{Edd}$. With an Eddington luminosity $L_\mathrm{Edd} = 4\pi G \mbh c / \kappa_\mathrm{es} = 9\times 10^5 L_\odot$ for the $25M_\odot$ BH, 
$\dot{M}_\mathrm{Edd} = L_\mathrm{Edd}/(\eta c^2) = 4\pi G \mbh c / (\eta\kappa_\mathrm{es})\sim 6 \times 10^{-7} M_\odot \mathrm{yr}^{-1},$
where here we have assumed an electron-scattering opacity of $\kappa_\mathrm{es}= 0.2(1+X)\,\mathrm{cm^2/g}$ for a fully ionized medium with hydrogen mass fraction $X=0.7$ and accretion efficiency of $\eta = 0.1$. As such, the BH accretion feedback is unlikely to have a significant effect on the total mass in the Hill sphere. However, the feedback may be energetically significant. For the power-laws represented as Equations~\ref{eq:ss_rho}-\ref{eq:ss_P}, the thermal energy of the system is bound only if some inner radius is specified. 
\begin{equation}
    K_\mathrm{H} = \int_{r_\mathrm{min}}^{\rhs} \frac{P}{\gamma - 1}4\pi r^2\mathrm{d}r = \frac{G\mbh m_\mathrm{H,s}}{2\rhs}\ln \left(\frac{\rhs}{r_\mathrm{min}}\right)
\end{equation}
One can consider $\tau_\mathrm{Edd}$, the timescale for Eddington limited feedback to alter the thermal energy in the Hill sphere
\begin{equation}
    \tau_\mathrm{Edd} = \frac{K_\mathrm{H}}{L_\mathrm{Edd}} = \frac{\sigma_T\left(1+X\right)m_\mathrm{H,s}}{16\pi cm_p \rhs}\ln \left(\frac{\rhs}{r_\mathrm{min}}\right)
\end{equation}
In the hydrodynamic simulations, this inner radius is the smoothing length $h=0.025\rhs$ and $\tau_\mathrm{Edd} \sim 1$yr, implying that Eddington limited energetic feedback could be significant. However, if the inner radius is much smaller, this timescale will increase and the feedback may become negligible.
 More significant changes may be expected from the inclusion of radiation pressure, which is neglected here for both the analytic modelling and the hydrodynamic simulation, or effects associated with stellar evolution such as nuclear fusion.

\subsubsection{Cyclonic Wind Morphology}
\label{sec:flow}
\begin{figure*}
    \includegraphics[width=2\columnwidth]{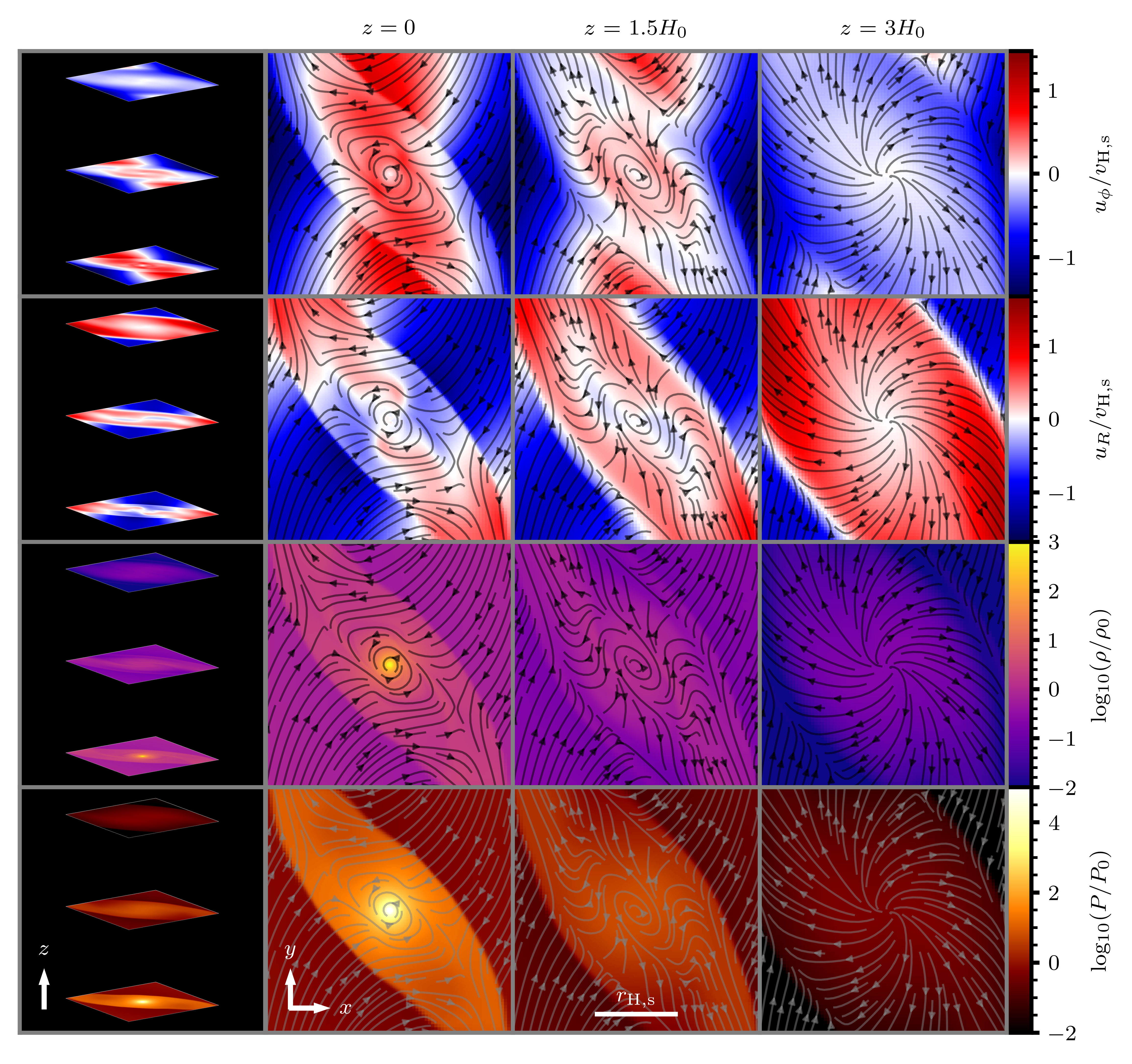}
    \caption{Slices in the $x$-$y$ disk plane for the fiducial model, cut at $z = [0,1.5H_0,3H_0]$. From the top, each row depicts the azimuthal gas velocity, radial gas velocity, gas density and gas pressure respectively. All velocities are in the BH frame. The minidisc at $z=0$ is limited to $r \lessapprox 0.5\rhs$: it hosts the hottest, densest gas and spins in prograde ($u_\phi > 0$). The minidisc drives a wind which expands to around $r\simeq 2\rhs$ by $z=3H_0$. This wind is less dense than the minidisc, but substantially denser than the ambient density at this height. By $z=3H_0$ the wind rotates in retrograde ($u_\phi < 0$) despite originating from a prograde minidisc; this is due to the cyclonic action of Coriolis force (see Equation~\ref{eq:cyclone}). At intermediate altitude, the central wind is still prograde, but the outer wind has been torqued to retrograde. In the leftmost panel the separations between planes have been increased by a factor 3 for illustrative purposes.}
    \label{fig:flow}
\end{figure*}

Figure~\ref{fig:flow} depicts the flow morphology about the BH at $t \sim 5$yr, with slices in the $x$-$y$ plane taken in and above/below the midplane. In the AGN midplane ($z=0$), the flow structure looks similar to that of 2D simulations. Gas is fed into the minidisc from the NE-SW quadrants (on horse-shoe orbits) resulting in a minidisc that spins in prograde ($u_\phi > 0$), despite the general sense of the flow in the shearing frame being retrograde ($u_\phi < 0$). The flow morphology changes when our view shifts to above the midplane at $z=3H_0$. Here the wind dominates the flow, with gas radially outflowing in all directions. The prograde rotation of the inner minidisc has been lost and the wind rotates in retrograde. Wind originating from the minidisc experiences Coriolis torque which drives it towards retrograde rotation. The effect of this force is most obvious when considering its action on the azimuthal component of the flow about the BH. It is straightforward to show from Equation~\eqref{eq:a_SMBH} that
\begin{equation}
    \label{eq:cyclone}
    \left(\frac{\mathrm{D}u_\phi}{\mathrm{D}t}\right)_{\mathrm{Coriolis}} = -2\Omega_0 u_R\,.
\end{equation}
This is cyclonic acceleration; inflowing components are driven toward prograde rotation ($u_R < 0 \implies \mathrm{D}_t u_\phi > 0$), whereas outflowing components are driven towards retrograde rotation ($u_R > 0 \implies \mathrm{D}_t u_\phi < 0$). This phenomena is identical to terrestrial cyclones in the Northern hemisphere: cyclones form around low pressure (inflow) regions and rotate anti-clockwise (prograde), whereas anticyclones form around high pressure (outflow) regions and rotate clockwise (retrograde). The minidisc behaves as a cyclone, whereas the wind behaves as an anticyclone. At intermediate heights between the minidisc-dominated and wind-dominated regimes lies a mixed region. In the central column of Figure~\ref{fig:flow} we can see that at $z=1.5H_0$ the streamlines near the BH are prograde, but moving away from the BH they invert to retrograde. As the height above the minidisc increases, all prograde components are lost and the outflow is dominantly retrograde. A wind that counter-rotates with respect to its minidisc has interesting implications for the circum-BH magnetic field structure, a point which is explored further in Section~\ref{sec:mag_cyc}.

\subsection{Binary Interactions}
\label{sec:close_enc}

We now consider the evolution of the BHs as they pass through a close encounter. During this close encounter, the fiducial binary loses sufficient orbital energy through gas gravity dissipation to form a stable binary. When discussing binary energies, we define $\Ebin$ in the binary centre-of-mass frame as
\begin{equation}
    \Ebin = \frac{1}{2}\mu |\vb{v}_1 - \vb{v}_2|^2 - \frac{G\mbin \mu}{|\vb{r}_1 - \vb{r}_2|},
\end{equation}
where $\mbin = m_1 + m_2$ and $\mu = \frac{m_1 m _2}{\mbin}$ are the total and reduced binary mass respectively. If $\Ebin < 0$, this gives a form for the semi-major axis $a$ as
\begin{equation}
    \Ebin = - \frac{G\mbin \mu}{2a}.
\end{equation}
The exact boundary for stability in hierarchical triple systems is dependent on a variety of different orbital parameters \citep{Mardling_2001, Tory_2022, Vynatheya_2022}. In this study we adopt a homogeneous threshold for all binaries, where a binary is considered stable to ionisation by the SMBH if $\Ebin < -\chi \Eh$ where $\chi = 2$ is a stability parameter derived empirically from simulations in \citet{Tory_2022} and $\Eh$ is the Hill energy, the characteristic energy scale of binaries in the shearing box
\begin{equation}
    \Eh = \frac{G\mbin \mu}{2\rh}.
\end{equation}
This energy is equivalent to the (absolute) energy of a binary with a semi-major axis $a = \rh$. With $\chi=2$, this criterion is equivalent to assigning stability to all binaries (with $e < 1$) that have instantaneous apoapsides $r_a = a(1 + e) < r_\mathrm{H}$. 

When discussing work done on the binary by either the SMBH or by the gravity of the gas, we report dissipation as the time derivative of the specific binary energy
\begin{equation}
    \label{eq:epsilon}
    \epsilon = \frac{\mathrm{d}}{\mathrm{d}t}\left(\frac{\Ebin}{\mu}\right) = \left(\vb{v}_1 - \vb{v}_2\right) \cdot \left(\vb{a}_1 - \vb{a}_2\right),
\end{equation}
where $\vb{a}_1$ and $\vb{a}_2$ are the accelerations acting on each binary component. When cumulative energy changes are reported, we integrate over the dissipation from specific sources e.g.
\begin{equation}
    \Delta E_\mathrm{gas} = \mu\int \epsilon_\mathrm{gas} \mathrm{d}t = \mu\int \left(\vb{v}_1 - \vb{v}_2\right) \cdot \left(\vb{a}_{\mathrm{gas},1} - \vb{a}_{\mathrm{gas},2}\right) \mathrm{d}t.
\end{equation}
and similarly for SMBH forces replacing the ``gas'' label with ``SMBH''. Here $\vb{a}_{\mathrm{gas},i}$ is the gravitational acceleration on a BH summed over all $N_c$ gas cells in the simulation
\begin{equation}
    \vb{a}_{\mathrm{gas},i} = \sum_{n=1}^{N_c} m_\mathrm{cell} \, g\left(\frac{\vb{r}_i-\vb{r}_n}{h}\right),
\end{equation}
where $m_\mathrm{cell}$ is the gas mass within an individual cell.

\subsubsection{Dissipation and Binary Formation}
\label{sec:disp}
\begin{figure*}
    \includegraphics[width=2\columnwidth]{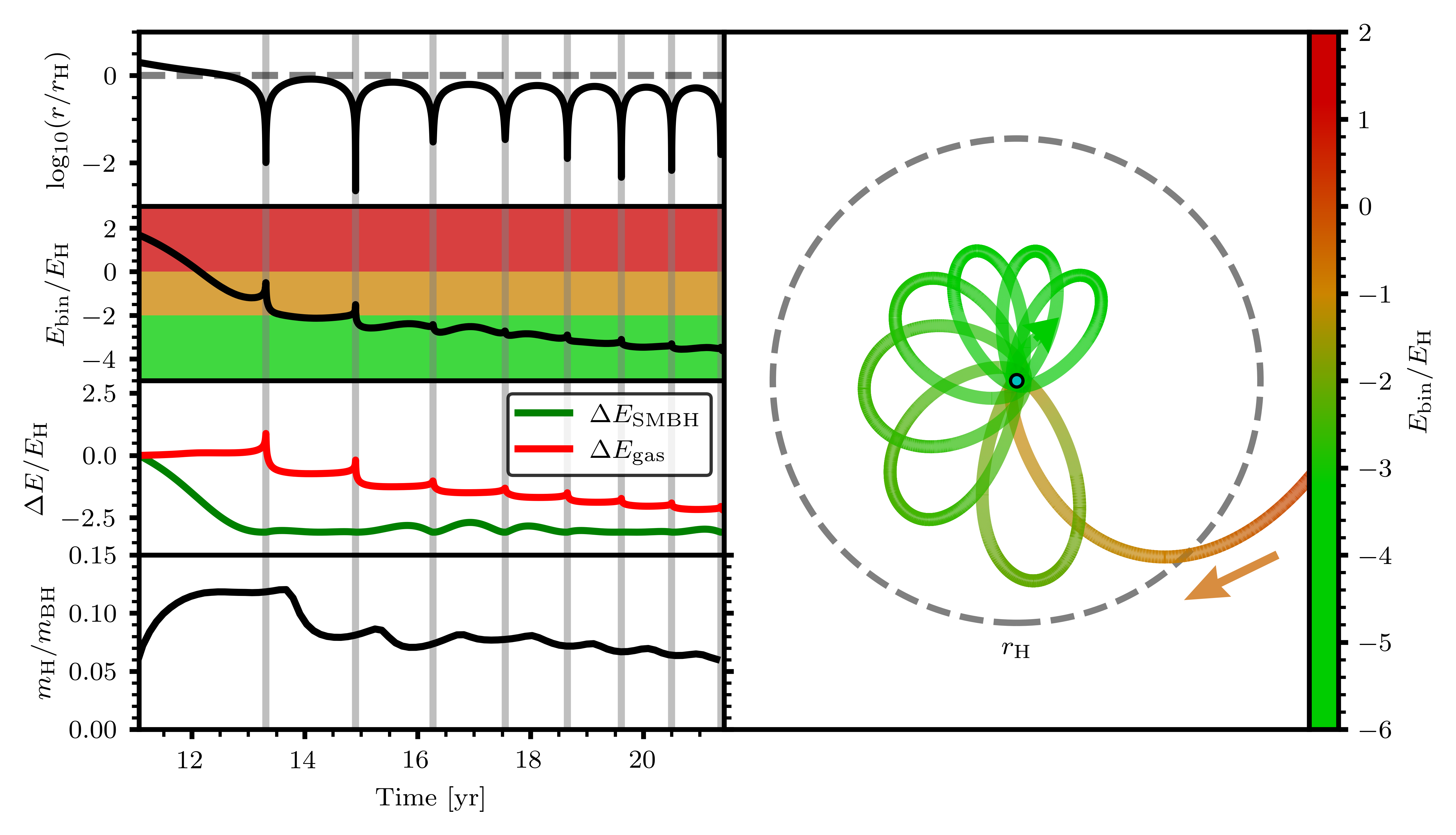}
    \caption{Formation of the fiducial binary, with panels on the left for the binary separation $r$, energy $\Ebin$, cumulative dissipation $\Delta E$ and gas mass within the mutual Hill sphere $m_\mathrm{H}$. In the right panel, the trajectory of the outer BH centred on the inner BH coloured by the instantaneous binary energy. During each periapsis passage (marked by vertical gray lines) the binary loses energy to gas gravity dissipation. Dissipation during the first periapsis is sufficient to form a stable binary, with further dissipation during subsequent periapsides continuing to harden the binary. Each periapsis is followed by a reduction in the Hill mass as detonations generated from minidisc collisions drive gas from the Hill sphere.}
    \label{fig:spec}
\end{figure*}

Figure~\ref{fig:spec} depicts the orbital evolution of the fiducial binary before and after the BHs pass through first close encounter, with panels for the binary separation, energy, trajectory and Hill mass. As the BHs approach each other, centrifugal forces steadily remove energy from the binary but the most important dissipation occurs immediately after first periapsis where the binary energy suddenly drops. The collision of the two minidiscs results in an excess of gas mass trailing the BH trajectories, exerting a drag by gas gravitation which removes energy from the binary. After a single close-encounter, the binary is stable against the SMBH, but relatively weakly bound. During the subsequent binary periapsis passages, there are further dissipative episodes which harden the binary. The chronology of the binary energetic evolution is identical to that observed in previous 2D hydrodynamical studies, with the binary energy dropping impulsively during each periapsis passage \citep{Li_2023, Rowan_2023, Rowan_2024, Whitehead_2024I, Whitehead_2024II}. Each periapsis passage is associated with a drop in the gas mass within the binary Hill sphere, $\mh$, as the novae generated at periapsis liberate mass from the Hill sphere (see Section~\ref{sec:novae}). However the mass reduction in 3D is not as dramatic as observed previously in 2D \citepalias{Whitehead_2024II}. This is because in 3D, the novae blasts are able to propagate away from the midplane, resulting in lesser disruption to the midplane where the majority of the gas mass lies. The first blast is the strongest, removing around a third of the gas from the Hill sphere, but subsequent blasts remove significantly less mass: after 6 periapsis passages, the Hill sphere has retained about half of its mass at first periapsis. We also document a decrease in hardening efficiency as the Hill mass drops: cumulatively, as much hardening occurs during the first two binary periods as during latter five. At the end of the simulation after 21.5yr, the final result is a binary with a semimajor axis of $a = 0.23\rh \sim 10^7 r_g$ and eccentricity $e\sim0.93$. While this binary is stable, its merger time as predicted by \citet{Peters_1964} is still $\sim 10^3$Gyr. Further hardening by gas will be required if the binary is to merge in a Hubble time, such evolution is not considered in this work. The long-term evolution of embedded binaries remains an active field of research \citep{Baruteau_2011, LiLai_evo, LiLai_eos, LiLai_visc, Dittmann_2024, Calcino_2024, Mishra_2024}.

\subsubsection{Disc Novae}
\label{sec:novae}
\begin{figure*}
    \includegraphics[width=2\columnwidth]{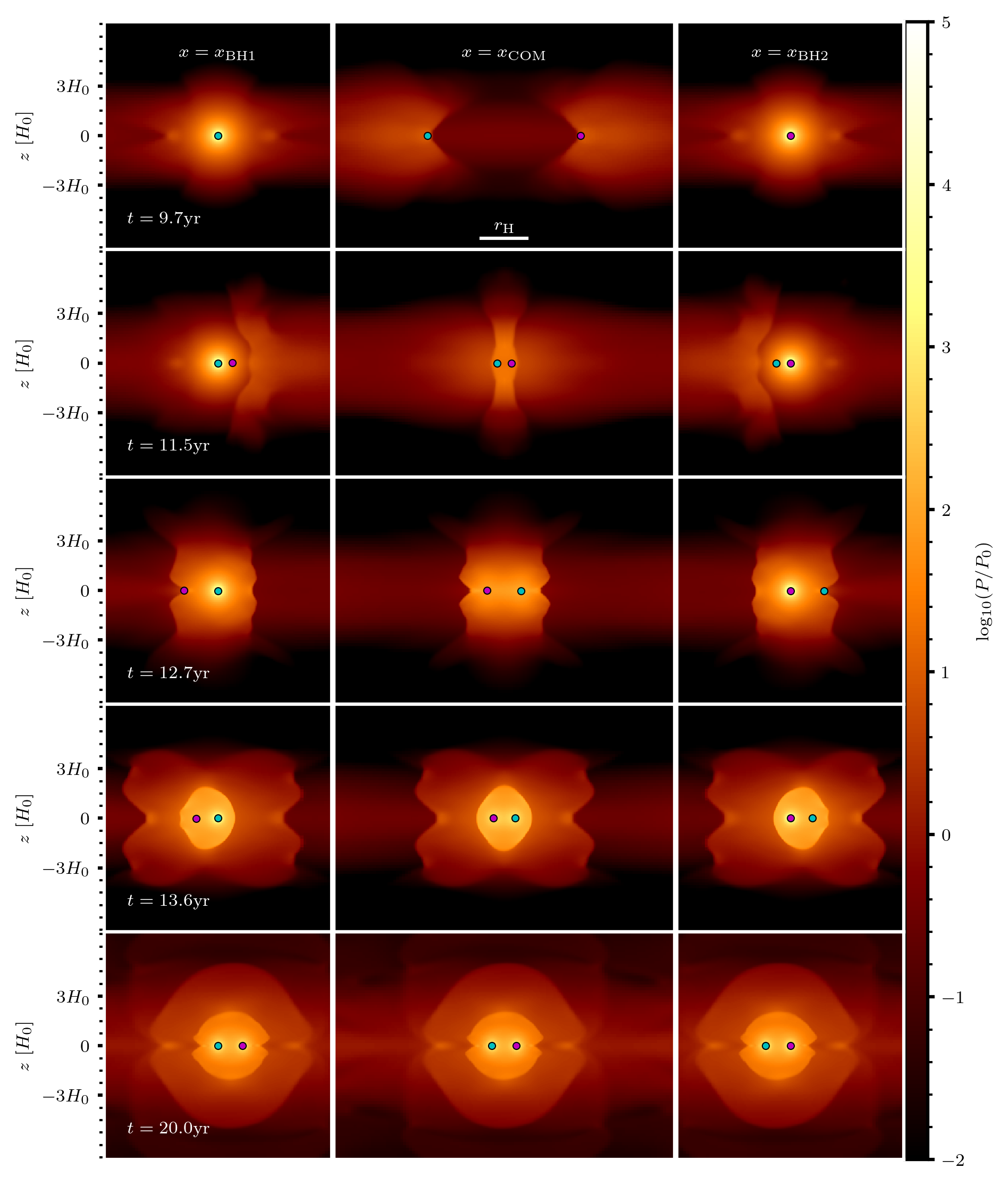}
    \caption{Logarithmic pressure profiles for the fiducial simulation sliced in the $y$-$z$ plane, with each row depicting a different epoch and each column centred on either binary component (in cyan and magenta) or the binary COM. \textit{First Row:} the gas structure before close encounter, showing isolated minidisc winds. \textit{Second Row:} as the BHs approach periapsis, their winds collide, forming a hot column of gas between the two. \textit{Third Row:} minidisc collision at periapsis drives strong shock heating. The shock escapes preferentially away from the midplane. \textit{Fourth Row:} following second periapsis, a new shock is generated. This shock is more regular due to the ejection of mass from the Hill sphere by the first shock. \textit{Fifth Row:} at late times the reduced binary period results in frequent shock generation. The shocks tend not to propagate in the midplane and are vertically compressed by the SMBH gravity leading to an spheroidal shape.}
    \label{fig:static_collision}
\end{figure*}

A key motivation for this study was to follow up the 2D results of \citetalias{Whitehead_2024II}, which observed strong blast waves generated during BH-BH close encounters due to minidisc collisions termed ``disc novae''. We observe these same blast waves being generated here in the fiducial simulation during first periapsis, with a hot fireball of gas generated around $t=12.5$yr from the inner minidiscs colliding. This can be seen in the lower panels of Figure~\ref{fig:static_collision}: at $t=12.7$yr the nova is visible as an expanding hot column of gas. The blast preferentially propagates in the vertical direction: due to the reduced gas mass away from the midplane this is the path of least resistance. The novae propagate fast enough to escape the Hill sphere before the next close encounter occurs. Each following periapsis passage ignites a new nova, by $t=20$yr the novae are more regular and propagate outwards as oblate ellipsoids. The novae are kinked in the midplane due to the increased gas density there restricting propagation. While the upper AGN atmosphere has already been significantly heated by the individual minidisc winds, the periapsis novae are able to eject hotter gas multiple scale heights above the midplane and so are likely to make significant contributions to the local thermal emission. The ability for the periapsis generated disc novae to escape away from the midplane instead of being forced to propagation only radially (as was the case in the 2D simulations of \citetalias{Whitehead_2024II}) means that the mass stripping action of the novae is suppressed. While each periapsis passages features a reduction in the Hill mass, the effect is less severe than in 2D and the resulting circum-binary environment is more massive. For comparison, in the fiducial binary of \citetalias{Whitehead_2024II}, disc novae were able to eject >80\% of the Hill mass within 2yr. Greater Hill mass retention has the potential to allow for more significant gas hardening post binary-formation, but this later evolution is not studied in this work.

\section{Parameter Study}
\label{sec:param}

Here we consider the full suite of 135 simulations across the 9 different AGN environments generated from varying the radial location in the AGN disc $R_0\in \left[5\times 10^3, 10^4, 2 \times 10^4\right]R_g$ and Eddington fraction of the AGN disc $l_E \in \left[0.05, 0.16, 0.50\right]$. In each environment we simulate 15 binary encounters with impact parameters $b \in [1.7,2.4]\rh$. We begin with a short discussion of the gas-free trajectories and then present the results from the full hydrodynamic simulations.

\subsection{Gas-free interactions}
\label{sec:gas-free}
\begin{figure*}
    \includegraphics[width=2\columnwidth]{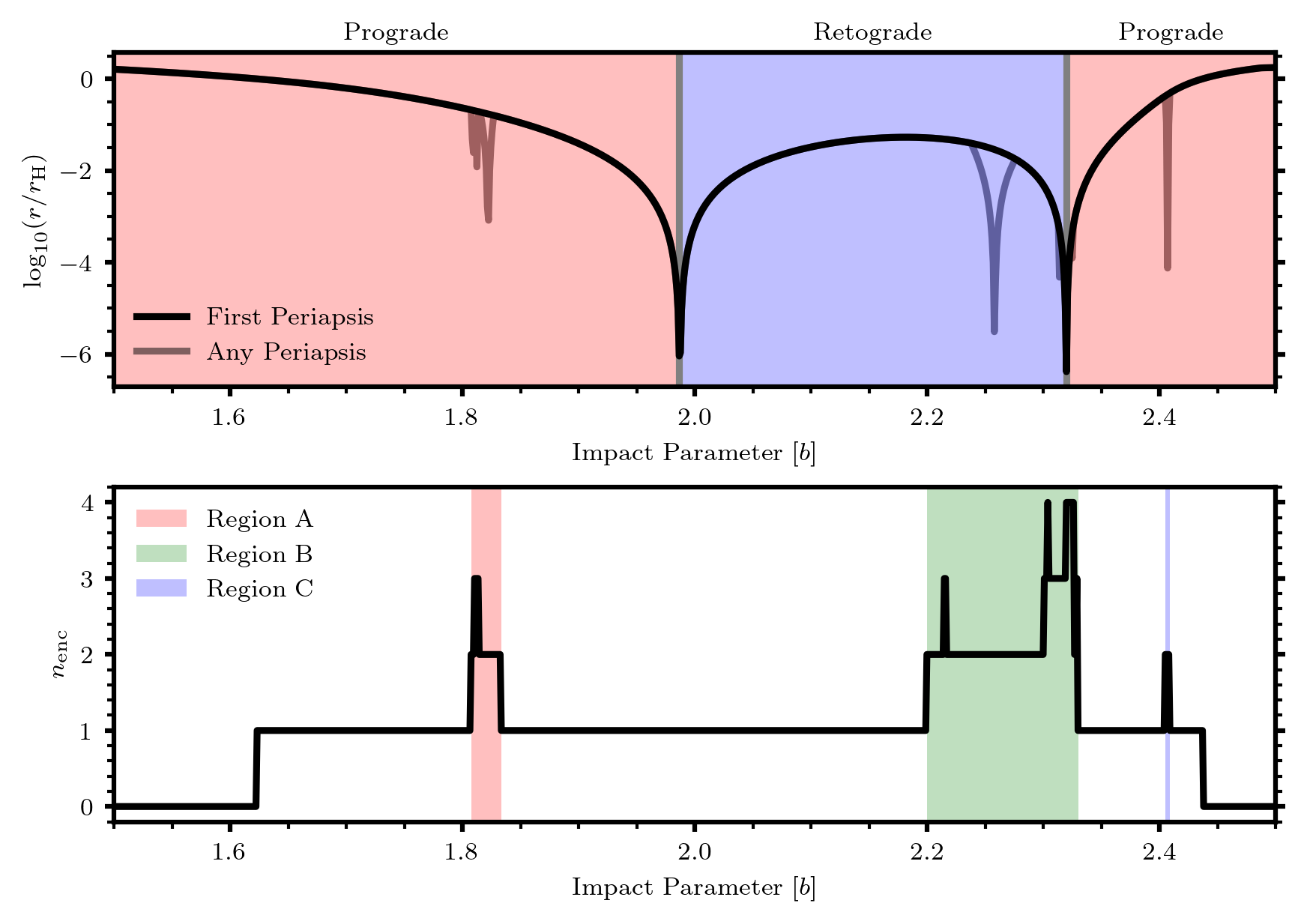}
    \caption{Spectrum of periapsis depths and number of encounters for gas-free interactions as the impact parameter $b$ is varied, generated from 1000 $n$-body simulations. Binaries with $b \sim 2.0, 2.3\rh$ will have very deep first close encounters driven by the SMBH torquing the binary angular momentum very close to zero before first periapsis. Binaries with impact parameters in the Jacobi Regions A, B and C will undergo multiple encounters before separating. Initial encounters can either be prograde or retrograde w.r.t the global AGN disc, separated by the zero angular momentum trajectories at $b \sim 2.0, 2.3\rh$}
    \label{fig:jacobi}
\end{figure*}

To better understand the spectrum of behaviours with gas, we briefly summarise the variety of encounters possible in gas-free interactions. Of specific interest are the Jacobi regions where the SMBH can drive multiple encounters between the BHs even when gas is not present \citep{Boekholt_2022}. Figure~\ref{fig:jacobi} gives a spectrum of close-encounter depths versus impact parameter for a gas-less system with the same initial conditions as our study. In the upper panel, we note the twin dips at $b \simeq 2.0, 2.3\rh$, these are the impact parameters associated with the deepest first periapsides. In these dips, the SMBH extracts the exact amount of angular momentum from the binary pre-encounter such that the binary angular momentum at periapsis is close to zero (resulting in a very small periapsis). These dips are prime regions for binary formation due to their significant periapsis depth, which our previous studies \citetalias{Whitehead_2024I} and \citet{Rowan_2024} found to be a key determinant for dissipation efficiency. In the bottom panel, we record the number of encounters $n_\text{enc}$ (defined as number of periapsides with $r_p < \rh$). There exist 3 regions where $n_\text{enc} > 1$ without an external dissipation mechanism, labelled $A$, $B$, $C$\footnote{Each of these regions contains a further 3 distinct sub-regions and so on, resulting in a fractal structure \citep{Boekholt_2022}}. These are Jacobi regions, which are also ideal for binary formation as multiple encounters allow for multiple dissipation events which cumulatively may drive stable binary formation. When gas is introduced, gas gravity acting on the BHs results in deflections from these gas-free trajectories, but this plot acts as a good introduction to the regions of interest in $b$ space.

\subsection{Resultant Binary Outcome}
\label{sec:outcomes}
We now consider the resultant state for the full suite of hydrodynamic simulations. We differentiate between outcomes by considering the number of orbits before binary formation can be determined as successful or unsuccessful. Binary formation is deemed successful if the binary energy $\Ebin < -2\Eh$. Encounters are defined as periapsides within the mutual Hill sphere. 
\begin{itemize}
    \item Type 1a: Unbound after single or no encounter. A single inefficient dissipation period results in no binary formation, the binary components separate without additional encounters.
    \item Type 1b: Unbound after multiple encounters. Secondary encounters may be driven by the SMBH or by perturbations by gas, but cumulatively dissipation is unable to stabilise the binary before the SMBH ionises it.
    \item Type 2a: Bound after single encounter. Following a single periapsis passage, the BHs are able to dissipate enough energy by gas gravity to form a stable binary.
    \item Type 2b: Bound after multiple encounters. While unstable after the first encounter, further cumulative dissipation over later periapsides is sufficient to stabilise the binary.
\end{itemize}

\begin{figure}
    \includegraphics[width=1\columnwidth]{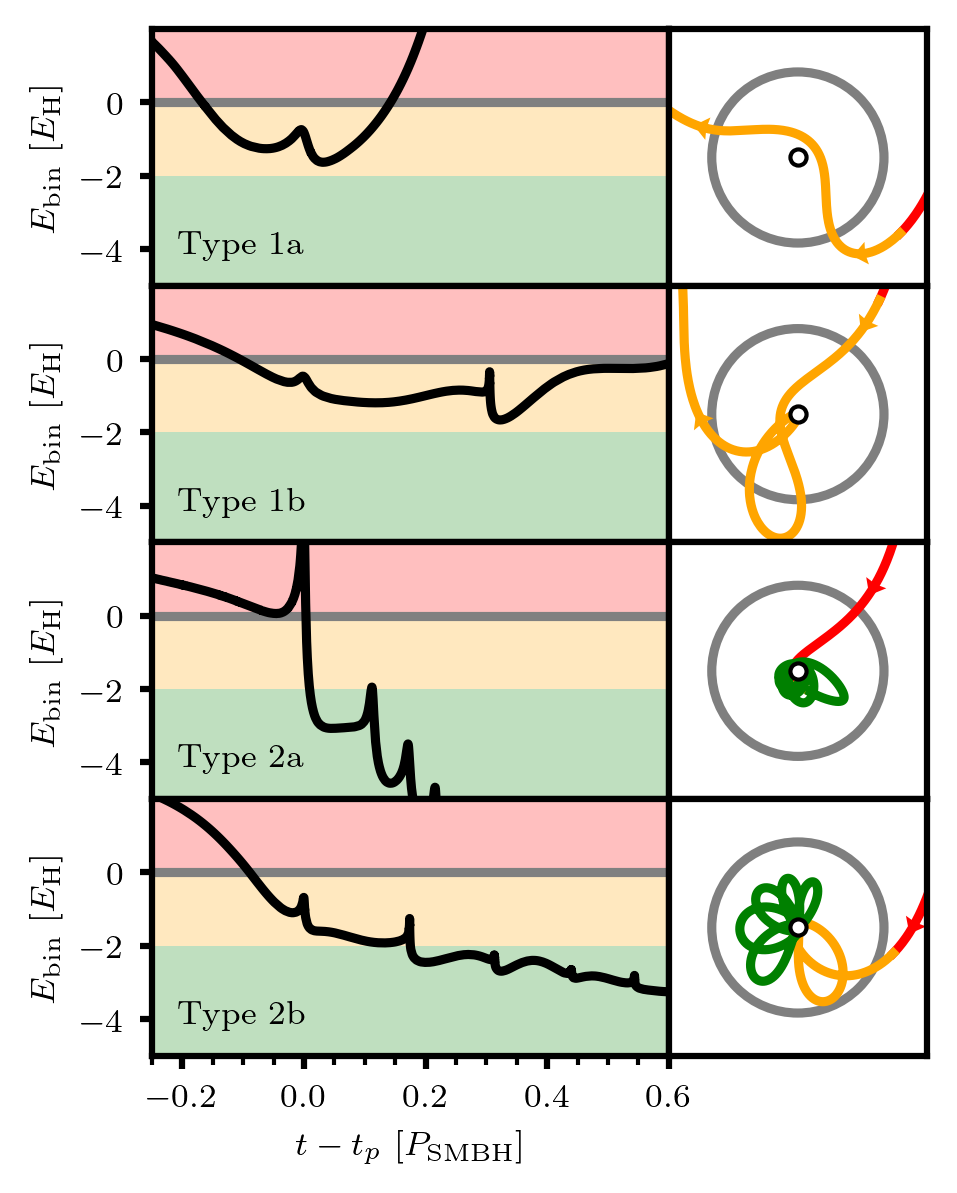}
    \caption{Four simulation runs depicting the range of outcome types (Type 1 for failure, Type 2 for success), where the colours red, orange and green represent unbound, quasi-bound and stably bound energetic regimes. While most runs attain $\Ebin < 0$ for some duration of their run, only runs with significant cumulative gas dissipation are able to cross the stability threshold of $\Ebin < -2\Eh$. For some runs, this threshold is only crossed after multiple periapsides, here the SMBH may assist binary formation by allowing for multiple encounters before ionisation (Type b encounters).}
    \label{fig:outcome_types}
\end{figure}

Figure~\ref{fig:outcome_types} depicts examples for the four outcome classes. The outcome of each flyby is dependent on both the BH trajectory (which will dictate the depth of encounter and the possibility for multiple Jacobi encounters) and the system's hydrodynamic state (as more massive minidiscs exert stronger acceleration by gas gravity). The relationship between the initial AGN environment and the resultant hydrodynamic state at first periapsis is discussed further in Section~\ref{sec:env_var}.

\begin{figure*}
    \includegraphics[width=2\columnwidth]{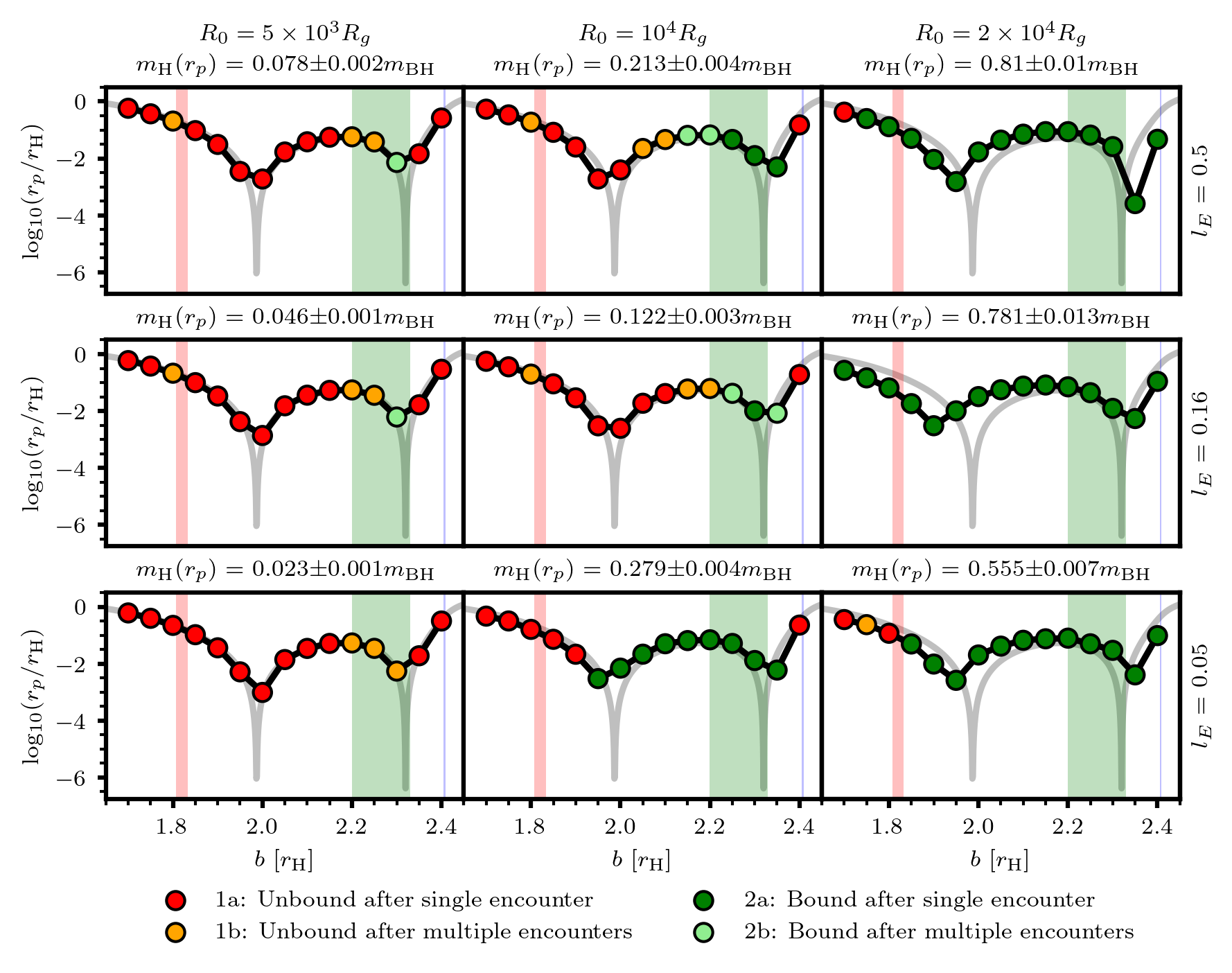}
    \caption{Binary formation outcomes for all 135 simulations in the suite, showing the separation at first periapsis $r_p$ as a function of the initial impact parameter $b$. Each sub-figure represents a unique AGN environment, generated by varying global Eddington fraction $l_E$ and distance from the SMBH $R_0$ by row and column respectively. The grey curve and vertical shaded regions show the gas-free depth of first periapsis and Jacobi regions from Figure~\ref{fig:jacobi}. Each environment is also labelled with the average Hill mass at first periapsis $\mh(\rp)$ : we note that generally, as $\mh(\rp)$ increases, so does the number of successful formation events (see also Figure~\ref{fig:mass}). This results in a strong bias towards more distant regions of the AGN disc, where the Hill radius is larger.}
    \label{fig:outcomes}
\end{figure*}

We document the final outcomes of all 135 hydro simulations in Figure~\ref{fig:outcomes} with the results of the gas-less simulations from Figure~\ref{fig:jacobi} shown in the background. The periapsis depths are generally similar to the gas-less trajectories, but show an inward skew for smaller impact parameters. We record also the gas mass in the mutual Hill sphere at first periapsis as $\mh(\rp)$; this is an important determinant in binary outcome as encounters with more local gas mass generally experience greater dissipation and so are more likely to form a binary. 

\subsubsection{Environmental Effects}
\label{sec:env_var}

\begin{figure}
    \includegraphics[width=\columnwidth]{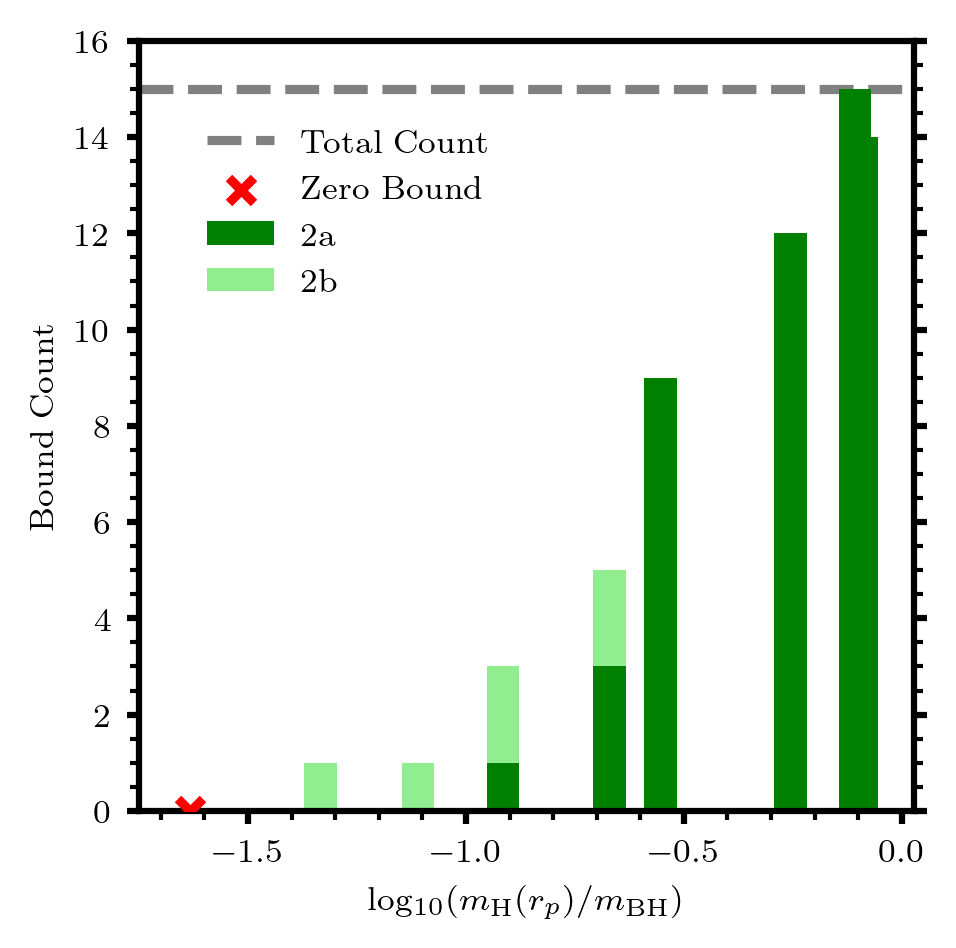}
    \caption{Hill mass at first periapsis, averaged over impact parameter $b$ for each of the 9 AGN environments, plotted against the number of successful binary formations. Successful binary formations are separated by type, ``2a'' if a stable binary is formed in a single flyby, or ``2b'' if stability is only achieved after multiple encounters. As $\mh(\rp)$ increases, so does the number of successful formation events; by $\mh(\rp) \sim 0.6 \mbh$ effectively all interactions in the simulated $b \in [1.7,2.4]\rh$ range result in binary formation. Conversely, below $\mh(\rp) \sim 0.1\mbh$ binary formation is rare and only possible via the Jacobi mechanism inducing multiple encounters. Below $\mh(\rp) \sim 0.03 \mbh$ we do not record any binary formation, though in principle it should be possible for suitably fine-tuned $b$.}
    \label{fig:mass}
\end{figure}

Figure~\ref{fig:outcomes} makes it clear that binary formation efficiency varies greatly across the 9 AGN environments analysed in this study. Comparing across the models, we break down the outcomes by type in Table~\ref{tab:env_outcomes}, including columns for the Hill masses in the ambient system $m_\mathrm{H,0}$, and at first periapsis $\mh(\rp)$. Figure~\ref{fig:mass} compares the number of successful binary formation events to $\mh(\rp)$. Generally we see that in environments with lower Hill masses, binary formation is more reliant on the Jacobi mechanism (Type b encounters), requiring multiple encounters to form stable binaries as each individual dissipation episode is too weak to ensure stability on its own. In very low Hill mass systems, such as documented in $(l_E, R) = (0.05, 5\times10^3 R_g)$, we record no binary formation. However, as the fractal structure of the Jacobi regions allows for an arbitrarily high number of encounters for suitably fine-tuned $b$ \citep{Boekholt_2022}, even seemingly insignificant gas masses could allow for stable binary formation over a suitably large number of encounters. However, as such interactions would only take place within such a narrow range of $b$-space, they are unlikely to contribute significantly to the overall binary formation rates. Conversely, in higher Hill mass environments where dissipation is stronger, the Jacobi mechanism becomes less important, as stable binaries are able to form from a single encounter. For the most massive systems in this study, binary formation is possible over the entire $b \in \left[1.7,2.4\right]\rh$ range. To leading order, to expect a capture likelihood of $(5\%, 20\%, 50\%, 100\%)$, we require Hill masses at periapsides of $\mh(\rp)\sim(0.05, 0.1, 0.25, 0.8)\mbh$.

\begin{figure}
    \includegraphics[width=\columnwidth]{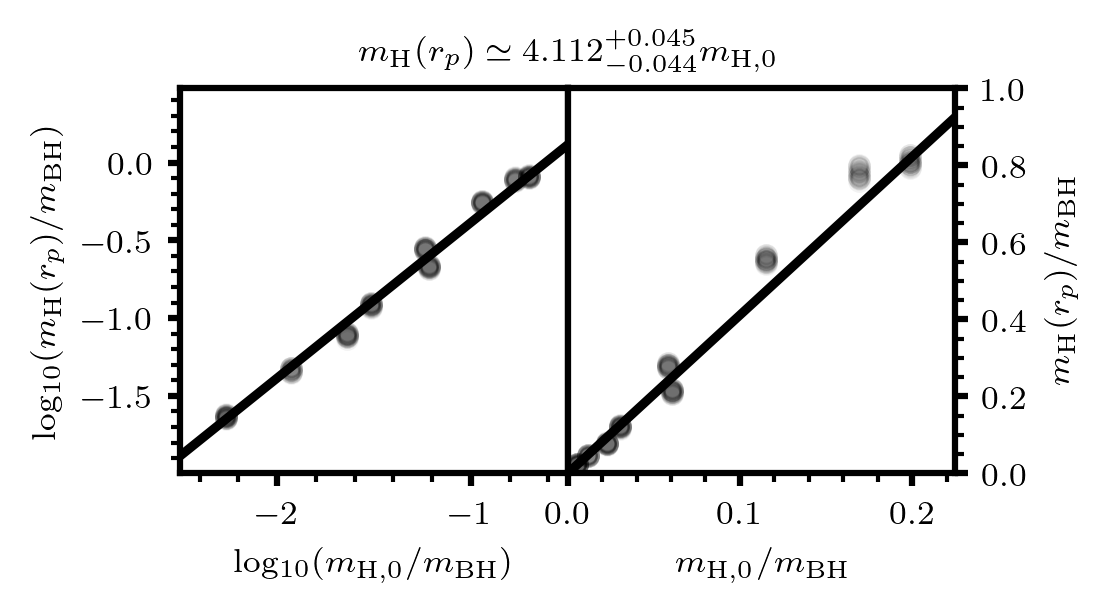}
    \caption{Ambient Hill mass $m_{\mathrm{H},0}$, against the Hill mass at first periapsis $\mh(\rp)$ for all simulations in the suite. We observe a strong linear relationship (fit in log space): more massive ambient environments result in more gas mass at the time of encounter. The gravitational attraction of gas onto both BHs before the first close encounter results in an enrichment of Hill mass to around 4 times the initial value.} 
    \label{fig:enrichment}
\end{figure}

Given the strong dependence of binary formation likelihood on the Hill mass at first periapsis presented in Figure~\ref{fig:outcomes}, we consider how readily this periapsis mass can be predicted from the ambient Hill mass. Figure~\ref{fig:enrichment} shows the relationship between the gas mass in the ambient Hill sphere (see Equation~\ref{eq:mh0}) and the Hill mass at first periapsis. We report a strong linear relationship between the two, with the mass at periapsis around 4 times the ambient mass for all of our simulations. As such, our findings support the notion that binary formation is typically most likely to be successful in regions where the ambient Hill mass is high, as this leads to more massive Hill masses at first periapsis, which in turn increases the efficiency of dissipation by gas and the likelihood of binary formation. As a rough approximation, capture likelihoods of $(5\%, 20\%, 50\%, 100\%)$, require ambient Hill masses of $m_\mathrm{H,0}\sim(0.01, 0.025, 0.063, 0.2)\mbh$.

\begin{table*}
    \centering
    \begin{tabular}{c c c c c c c c c}
    \hline
     $R_0/R_g$        & $l_E$   & $m_\mathrm{H,0}$  & $\mh(r_p) / \mbh$   & Type 1a   & Type 1b   & Type 2a   & Type 2b   & Total Bound \% \\ \hline
     $5\times 10^3$ & 0.05    & 0.005             & $0.023\pm0.001$     & 14        & 3         & 0         & 0         & 0\% \\
     $5\times 10^3$ & 0.16    & 0.011             & $0.046\pm0.001$     & 11        & 3         & 0         & 1         & 7\% \\
     $5\times 10^3$ & 0.50    & 0.023             & $0.078\pm0.002$     & 11        & 3         & 0         & 1         & 7\% \\
     $1\times 10^4$ & 0.05    & 0.058             & $0.279\pm0.004$     & 6         & 0         & 9         & 0         & 60\% \\
     $1\times 10^4$ & 0.16    & 0.030             & $0.122\pm0.003$     & 9         & 3         & 1         & 2         & 20\% \\ 
     $1\times 10^4$ & 0.50    & 0.061             & $0.213\pm0.004$     & 7         & 3         & 3         & 2         & 33\% \\ 
     $2\times 10^4$ & 0.05    & 0.115             & $0.555\pm0.007$     & 2         & 1         & 12        & 0         & 80\% \\
     $2\times 10^4$ & 0.16    & 0.169             & $0.781\pm0.013$     & 0         & 0         & 15        & 0         & 100\% \\ 
     $2\times 10^4$ & 0.50    & 0.200             & $0.810\pm0.010$     & 1         & 0         & 14        & 0         & 93\% \\ [1ex]
    \hline
    \end{tabular}
    \label{tab:env_outcomes}
    \caption{Summary of outcome variation across AGN environments, here $m_\mathrm{H,0}$ and $\mh(r_p)$ are the gas masses in the mutual Hill sphere at $t=0$ and at first periapsis respectively. Outcomes types are as defined in Section~\ref{sec:outcomes}. As $\mh(r_p)$ increases, we generally observe an increase in binaries formed, especially during the first encounter. For a clearer trend between formation likelihood and $\mh(\rp)$, see Figure~\ref{fig:mass}.}
\end{table*}

\subsubsection{Trajectory Dependence}
\label{sec:traj}
In Section~\ref{sec:env_var}, we showed that systems with higher Hill masses exhibited a greater fraction of successful binary formations. However, as seen in Figure~\ref{fig:outcomes}, there is dependence also on $b$, the impact parameter between interacting BHs. In the highest mass systems, this dependence is less significant, with binary formation successful for all runs experiencing a suitable close first encounter. However for intermediate and lower mass systems, the initial impact parameter is the determining factor between binary formation and ionisation. We observe a strong bias towards formation in the deep periapsis valley around $b\sim2.3\rh$, most notable in the $(l_E,R) = (0.16,10^4 R_g)$ and $(0.50, 10^4R_g)$ systems, for which $\mh(\rp) \sim 0.12\mbh$ and $0.22\mbh$ respectively. In these simulations, binary formation is only successful near Jacobi region B, where the action of the SMBH induces not only a deep first periapsis, but allows for multiple encounters before ionisation (see Section~\ref{sec:gas-free}). These multiple encounters turn out to be the deciding factor; deep first encounters can be achieved for impact parameters close to either valley at $b=2.0,2.3\rh$, but it is the presence of the Jacobi region at $b=2.3\rh$ that increases the likelihood of binary formation. In/near the Jacobi regions, even if a binary is not stabilised during the first periapsis, dissipation during subsequent close encounters can harden the binary and stabilise it past the $\Ebin < -2\Eh$ threshold. One might then enquire why these Jacobi regions exist at these ranges in $b$ space. The key factor turns out to be the orientation of the binary following first periapsis; for systems with inefficient dissipation, if the semi-major axis is aligned with the $\hat{\bm{x}}$ direction (i.e aligned with the radial direction to the SMBH), ionisation is favoured, but if the semi-major axis is aligned with the $\hat{\bm{y}}$ direction (i.e. perpendicular to the radial direction to the SMBH), the system can undergo multiple encounters before ionisation.

\begin{figure*}
    \includegraphics[width=2\columnwidth]{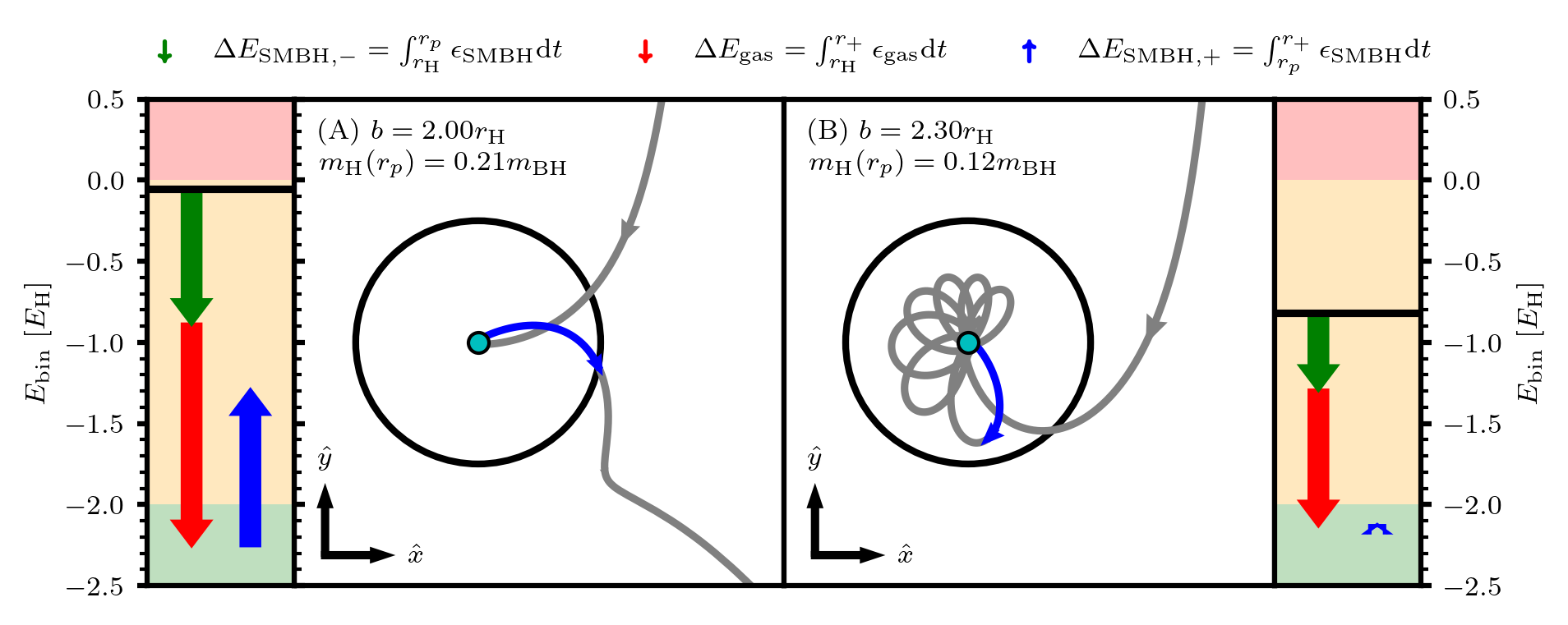}
    \caption{Comparison between simulations with different impact parameters and periapsis Hill masses. The central panels depict the trajectory of the outer BH in the inner BH frame, the outer panels show the energy evolution. Despite encounter (A) featuring a greater Hill mass and stronger dissipation at periapsis (seen in the larger red arrow), it fails to form a binary, whereas encounter (B) succeeds. While both systems reach similar energies considering only the work done by gas and the SMBH on infall (red and green arrows), post periapsis system (A) features a trajectory closely aligned with the $\hat{\bm{x}}$ axis (i.e. the direction of the SMBH), resulting in a strong injection of energy by centrifugal action (blue arrow). Conversely, system (B) features a relatively wide first period which is perpendicular to the $\hat{\bm{x}}$ axis and so the work done by the SMBH between periapsis and apoapsis is effectively zero. As a result, system (A) ionises immediately but system (B) is able to stabilise over subsequent periods. In the centre panels, the blue arrows represent the integration periods for the post-periapsis SMBH dissipation $\Delta E_{\mathrm{SMBH},+}$.}
    \label{fig:arguments}
\end{figure*}

To better explain this dependence on the post-first-periapsis orientation (which in turn depends on the initial impact parameter), we introduce 3 energy changes pertinent to the binary outcome:
\begin{itemize}
    \item $ \Delta E_{\mathrm{SMBH,-}} = \mu\int_{\rh}^{r_p}\epsilon_\mathrm{SMBH}\mathrm{d}t$ is the work done by the SMBH as the binary approaches first periapsis from the Hill intersection. This work is most sensitive to the BH trajectory and independent of minidisc mass.
    \item $\Delta E_\mathrm{gas} = \mu\int_{\rh}^{r_+}\epsilon_\mathrm{gas}\mathrm{d}t$ is the work done by the gas between the Hill intersection and the first apoapsis (or exit of Hill sphere). This source of dissipation is generally stronger for deeper periapsides, but even more importantly is stronger for systems with more massive minidiscs.
    \item $\Delta E_{\mathrm{SMBH,+}} = \mu\int_{r_p}^{r_+}\epsilon_\mathrm{SMBH}\mathrm{d}t$ is the work done by the SMBH as the binary orbits from periapsis to first apoapsis (or exit of Hill sphere). This work is a function of the post-periapsis trajectory, which is dependent on both the incident trajectory and the strength of gas dissipation during periapsis.
\end{itemize}
The energy of the binary after its first close encounter will depend on the energy upon entering the Hill sphere, and the sum of these three energies. Figure~\ref{fig:arguments} displays the energetic evolution for two simulations with disparate impact parameters and Hill masses. We can see that the key distinguishing feature between the two is the energy injected by the SMBH post-periapsis (shown in blue). While both systems attain similar energies when considering only the gas dissipation and SMBH dissipation on infall, system (A) features significant energy injection by the SMBH post-periapsis. This is due to the binary's alignment with the $\hat{\bm{x}}$ axis; as the SMBH's centrifugal acceleration also acts in the radial $\hat{\bm{x}}$ direction there is a positive alignment between the binary velocity and the acceleration which results in energy ejection (see Equation~\ref{eq:epsilon}). Conversely, in system (B) the binary components are more closely aligned with the $\hat{\bm{y}}$ axis during the first period and so the action of the centrifugal acceleration is negligible. Not only is the magnitude of the centrifugal action reduced when the binary is aligned with the $\hat{\bm{y}}$ axis, the binary velocity is now perpendicular to the centrifugal acceleration, resulting in minimal work being done on the binary. As such, system (B) is able to perform further encounters before ionisation; when sufficient gas is present, dissipation during these periods then stabilises the binary. See Appendix~\ref{sec:jac_arg} for further detail as to how the initial binary orientation can be used to identify the locations of all 3 Jacobi regions.

\subsubsection{Modelling Approximations}
\label{sec:model}
\begin{figure}
    \includegraphics[width=\columnwidth]{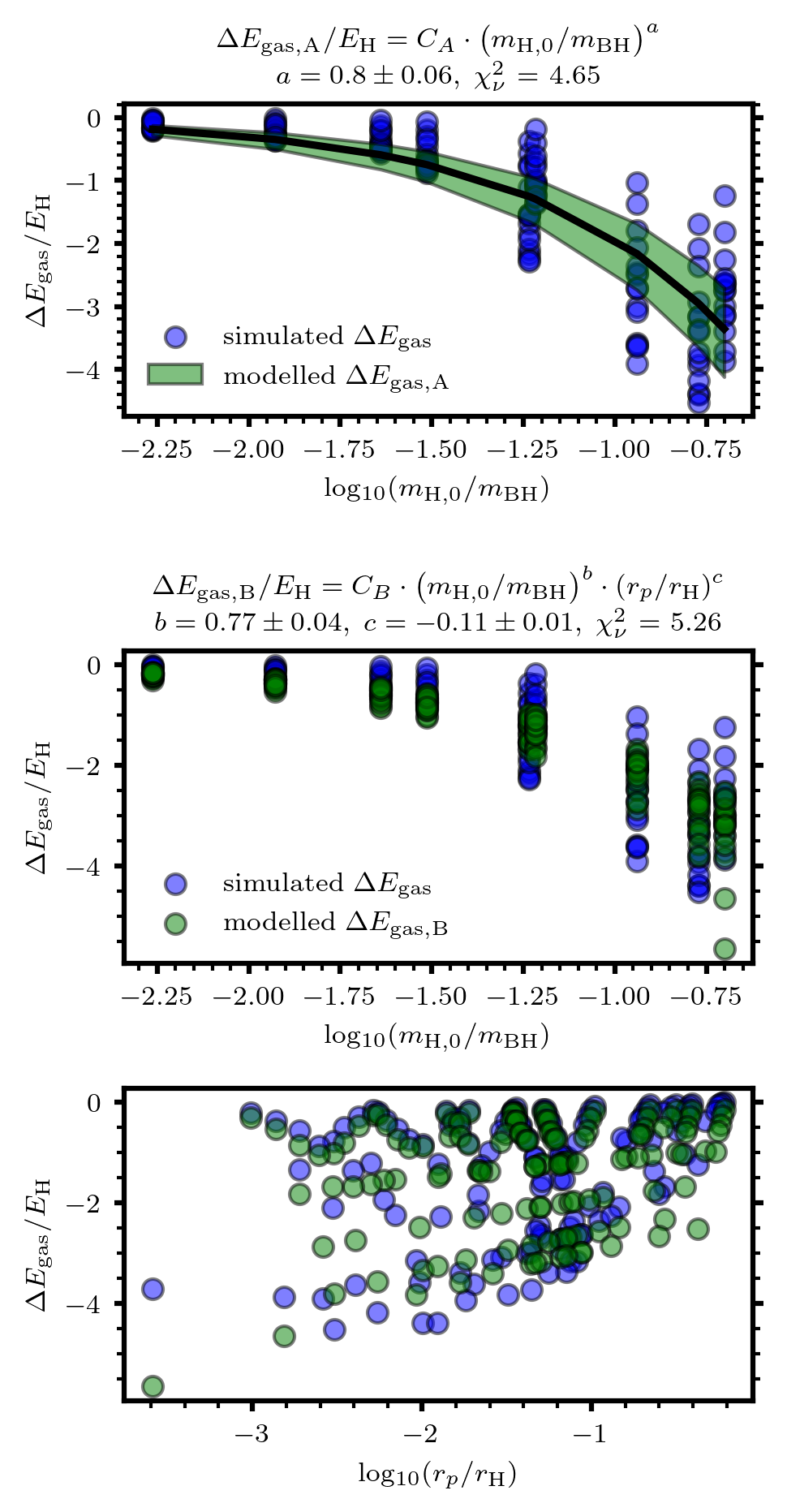}
    \caption{Predicted gas dissipations from Models A and B (Equations~\ref{eq:model_A} and ~\ref{eq:model_B}), compared against the true simulated dissipations. In the top panel, Model A considers only the effects of the ambient Hill mass $m_{\mathrm{H},0}$, giving one dissipation for each AGN environment. In the bottom two panels, Model B requires both $m_{\mathrm{H},0}$ and the depth of first periapsis $\rp$, giving a dissipation for each individual simulation. Both models show a significant scatter from the true results, but are able to reproduce the general trends reasonably well. Full fitting parameters for both models can be found in Table~\ref{tab:fit_param}.}
    \label{fig:disp_fit}
\end{figure}

Given the high cost associated with hydrodynamic simulation, we put forward fitting formulae and methodology to allow for inexpensive predictions of binary formation likelihood. As established in Sections~\ref{sec:env_var} and~\ref{sec:traj} the outcome of a potential binary formation event is sensitively dependent on both the ambient AGN conditions and the BH trajectories. While the ambient conditions should be readily available for a given disc prescription, the BH trajectories may not i.e. in Monte Carlo simulations. To account for this potential restriction, we provide two formulae for modelling the strength of dissipation by gas during a binary close encounter. The parameters for these formulae are fit directly from the 135 hydrodynamic simulations presented in this paper. 
\begin{equation}
    \label{eq:model_A}
    \mathrm{Model\;A:} \quad \frac{\Delta E_\mathrm{gas,A}}{\Eh} = C_A \cdot \left(\frac{m_{\mathrm{H},0}}{m_\mathrm{BH}}\right)^b 
\end{equation}
This model allows for predictions of the gas dissipation using only the ambient Hill mass $m_{\mathrm{H},0}$: a single quantity easily accessible to a wide variety of simulation types. 
\begin{equation}
    \label{eq:model_B}
    \mathrm{Model\;B:} \quad \frac{\Delta E_\mathrm{gas,B}}{\Eh} = C_B \cdot \left(\frac{m_{\mathrm{H},0}}{m_\mathrm{BH}}\right)^b \cdot \left(\frac{\rp}{\rh}\right)^c
\end{equation}
This model allows for extra additional information to be used if the depth of the close encounter $\rp$ is also known. Table~\ref{tab:fit_param} records the best-fitting parameter values and dispersions for both of these models. \citetalias{Whitehead_2024I} performed a similar analysis for 2D isothermal close encounters, finding that $\Delta E_\mathrm{gas} \propto \rho_0^\alpha r_p^\beta$, with $\alpha=1.01\pm0.04$ and $\beta=-0.43\pm0.03$ (see Table 1 of \citetalias{Whitehead_2024I}), with a similar constraint on $\beta$ from \cite{Rowan_2024}. As \citetalias{Whitehead_2024I} investigated a single distance $R_0$ from the AGN, $\rho_0$ in that study and $m_{\mathrm{H},0}$ here are directly proportional. For these 3D adiabatic encounters, we find the power law in gas mass to be slightly softened, with $a=0.80\pm0.06$ for Model A and $b=0.77\pm0.04$ for model B. The slope in $\rp$ is significantly reduced, with $c=-0.11\pm0.01$ for model B. The reduction in the dependence of dissipation on the periapsis depth is consistent with a reduction in compactness within the Hill sphere (see Figure~\ref{fig:cmdf}). A more extended distribution of gas mass around the BHs means that even relatively shallow periapsides can encounter significant gas mass. Figure~\ref{fig:disp_fit} compares the predicted dissipations of both models against the simulated dissipations for all simulations in the suite. For both models, we use the reduced chi-squared metric $\chi^2_\nu$ to compare the quality of the fits, calculating $\chi^2_{\nu,\mathrm{A}} = 4.65$ and $\chi^2_{\nu,\mathrm{A}} = 5.26$. Averaging over all models in the suite, we find root mean square errors of $\mathrm{RMS_A} = 0.62\Eh$ and $\mathrm{RMS_B} = 0.50\Eh$ (cf. $\Delta E_\mathrm{gas} < -2\Eh$ is necessary to definitively ensure capture). We do not find that Model B significantly outperforms Model A, and the quality of both fits are comparable, both providing reasonable estimates as to the gas dissipation. For a fixed value of $m_\mathrm{H,0}$, Model B should provide a more accurate model for $\Delta E_\mathrm{gas}$ over a range of $\rp$ values. There are likely other factors that contribute to the gas dissipation that we have not fit for here, such as the effects of disc temperature and the orientation of the BH trajectories. We suggest that models A and B are best used as statistical approximations for binary likelihood on a population level, such as within large Monte-Carlo simulations. 

\begin{table}
    \centering
    \begin{tabular}{c c c c c c}
    \hline
     Model & Parameter $v$  & $\mu_v$   & $\sigma_v$ & $\chi^2_\nu$ & RMS \\ \hline \hline
     A & $C_A$      & -12.28        & 1.44 & 4.65   & 0.62$\Eh$  \\
     &$a$           & 0.80          & 0.06 &        &  \\ \hline
     B &$C_B$       & -7.92         & 0.84 & 5.26   & 0.50$\Eh$    \\
     &$b$           & 0.77          & 0.04 &        &\\
     &$c$           & -0.11         & 0.01 &        & \\ [1ex]
    \hline
    \end{tabular}
    \caption{Best fit parameters $v$ (with means $\mu_v$ and standard deviations $\sigma_v$), reduced chi-squared $\chi^2_\nu$ and root mean square RMS for dissipation Models A and B (see Equations~\ref{eq:model_A} and~\ref{eq:model_B}). Despite Model B requiring more information, both models show similar performance in reproducing the magnitude of dissipation by gas during close encounters.}
    \label{tab:fit_param}
\end{table}

\begin{figure*}
    \includegraphics[width=2\columnwidth]{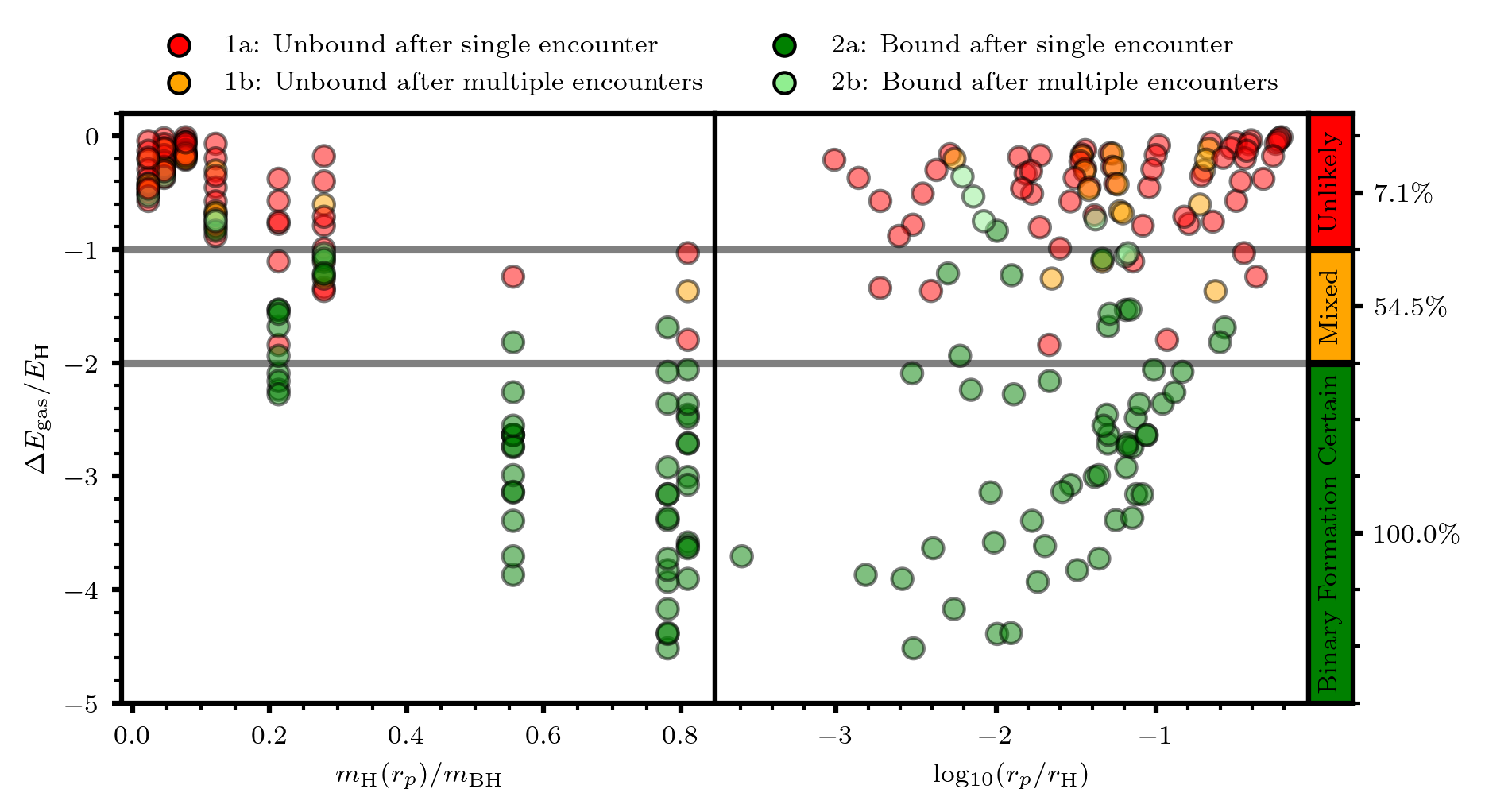}
    \caption{Energy dissipated to gas during the first close encounter $\Delta E_\mathrm{gas}$ for all simulations in the suite, plotted against the Hill mass at first periapsis $\mh(\rp)$ and depth of first periapsis $\rp$. All dissipations are coloured by the resultant binary outcome. Broadly, the outcome space can be separated into 3 regions: if $\Ebin < -2\Eh$ binary formation is very likely, if $\Ebin > -\Eh$ is it very unlikely and for intermediate dissipation strengths the outcome is dictated by the BH dynamics. The percentage of bound systems in each region is listed to the right of the plot. We see that while $\mh(\rp)$ is generally a better indicator of $\Ebin$ than $\rp$, there is significant spread in $\Ebin$ for simulations with similar $\mh(\rp)$.}
    \label{fig:scatter_disp}
\end{figure*}

As established in Section~\ref{sec:traj}, the magnitude of dissipation by gas is not always sufficient to determine the success of binary formation. Figure~\ref{fig:scatter_disp} depicts the relationship between the gas dissipation and the binary outcome, as a function of both the Hill mass $\mh(\rp)$ and first periapsis depth $\rp$. While it is generally the case that formation is favoured for systems experiencing significant dissipation ($\Delta E_\mathrm{gas} < -2\Eh$) and disfavoured for systems experiencing insufficient dissipation ($\Delta E_\mathrm{gas} > -\Eh)$, there are intermediate dissipation strengths where formation is determined by the BH trajectories and orientations. We can define 3 rough regions in the $\Delta E_\mathrm{gas}$ space:
\begin{itemize}
    \item $\Delta \Ebin > -\Eh$: Ionisation is very likely. A small fraction of systems are able to form stable binaries via the Jacobi mechanism.
    \item $-2\Eh < \Delta \Ebin \leq -\Eh$: Ionisation and formation are of similar likelihood. The exact outcome is dependent on the BH trajectory post periapsis (see Section~\ref{sec:traj}).
    \item $\Delta \Ebin \leq -2 \Eh$: All systems form stable binaries. Strong dissipation by gas results in a stable binary after a single close encounter. 
\end{itemize}
Using these approximate boundaries and the dissipation modelling formulae A and B, we propose a basic binary formation modelling description motivated by the statistics of our simulation suite. To determine whether or not a binary will form, consider: 
\ol{%
    \li{If the binary components are unable to reach a separation of $r<\rh$, no binary will be formed as no close encounter occurs.}
    \li{Given a close encounter occurs, use models A or B to predict the strength of dissipation $\Delta E_\mathrm{gas,pred}$.}
    \li{Determine the likelihood of binary formation from the magnitude of the gas dissipation:}
        \ol{%
            \li{If $\Delta E_\mathrm{gas,pred} > -\Eh$, assume that binary formation is unsuccessful. This results in a few rare Jacobi captures being discarded.}
            \li{If $-2\Eh < \Delta E_\mathrm{gas,pred} < \Eh$, assume a binary forms with 50\% likelihood.}
            \li{If $\Delta E_\mathrm{gas,pred} < -2\Eh$, assume that binary formation is successful.}
        }
}
Applying this predictive methodology to our simulations, we compare the simulated and predicted formation rates in Table~\ref{tab:model}. While Model B requires extra information in the form of the periapsis depth, Model A is able to better reproduce the simulated formation rates. Both models show deviation from the true rates, but are able to provide reasonable order-of-magnitude estimates as to the likelihood of binary formation. We hope that these models will aid Monte-Carlo simulations of BH interactions in AGN \citep{Tagawa_2020,McKernan_2024_I, Delfavero_2024, Rowan_2025a}.

\begin{table}
    \centering
    \begin{tabular}{c c c c c}
    \hline
     $R_0/R_g$        & $l_E$   & Simulated Rate  & Predicted Rate A   & Predicted Rate B \\ \hline
     $5\times 10^3$ & 0.05    & 0\%             & 0\%       & 3\%        \\
     $5\times 10^3$ & 0.16    & 7\%             & 0\%       & 0\%       \\
     $5\times 10^3$ & 0.50    & 7\%             & 0\%       & 0\%       \\
     $1\times 10^4$ & 0.05    & 60\%            & 50\%      & 23\%       \\
     $1\times 10^4$ & 0.16    & 20\%            & 0\%       & 7\%         \\ 
     $1\times 10^4$ & 0.50    & 33\%            & 50\%      & 23\%         \\ 
     $2\times 10^4$ & 0.05    & 80\%            & 100\%     & 93\%         \\
     $2\times 10^4$ & 0.16    & 100\%           & 100\%     & 83\%         \\ 
     $2\times 10^4$ & 0.50    & 93\%            & 100\%     & 60\%         \\ [1ex]
    \hline
    \end{tabular}
    \label{tab:model}
    \caption{Rate predictions for each of the 9 AGN environments, comparing between models A and B for gas dissipation (see Equations~\ref{eq:model_A} and~\ref{eq:model_B}). Despite the coarseness in its output (as the effect of BH trajectory is ignored), Model A provides a slightly better match to the simulated rates, with some over-estimation in the $R_0=2\times 10^4 R_g$ systems.}
\end{table}

\subsection{Observable Signatures}
\label{sec:observables}

\begin{figure}
    \includegraphics[width=\columnwidth]{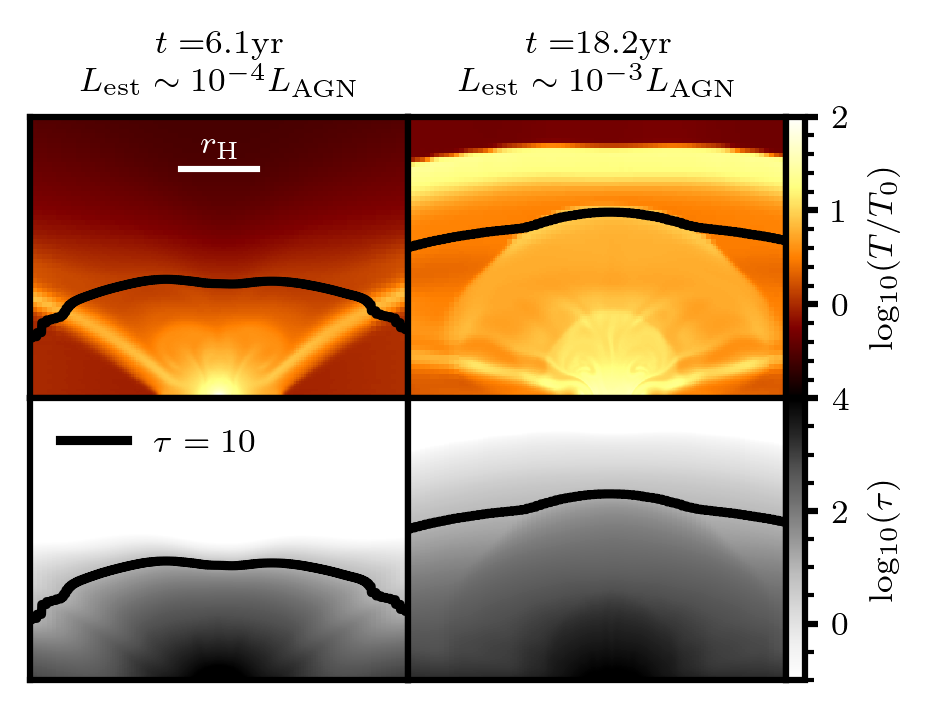}
    \caption{Gas temperatures $T$ and optical depths $\tau$ in the fiducial system, for a black hole in isolation (left) and post binary formation (right). The midplane is very optically deep ($\tau > 10^4$), but the upper wind and hot novae shells are significantly less obscured ($\tau \sim 10^2$).  High above the midplane lies a layer of artificially hot under-dense gas that would have cooled were radiative transfer included in the simulation. Over-layed in black is the $\tau > 10$ limit imposed for luminosity estimations $L_\mathrm{est}$. The system is an order of magnitude brighter post binary formation than during the wind phase.}
    \label{fig:tau}
\end{figure}

Unlike for our previous 2D study \citepalias{Whitehead_2024II}, we are unable to readily present observable signatures of embedded BHs in isolation or during close encounter in this study. This comes from the inability to account for gas cooling self-consistently in 3D without including radiative transport, which would significantly increase the computational expense. Previous approximations from the 2D system are no longer valid as vertical radiative equilibrium cannot be assumed (see Equations 18-23 from \citetalias{Whitehead_2024II}). Luminosities might be approximated post-simulation from the frequency-integrated radiative transfer solution,
\begin{equation}
    \label{eq:rad_trans}
    l = \int_{0}^{\tau_\mathrm{max}} \sigma_{\rm SB}T^4\,e^{-\tau}\mathrm{d}\tau = \int_{0}^{z_\mathrm{max}} \sigma_{\rm SB}T^4\,\kappa\rho e^{-\tau_z}\mathrm{d}z
\end{equation}
where $l$ is the thermal emission per unit area, $\sigma_{\rm SB}$ is the Stephan-Boltzmann constant, $\kappa$ is the absorption opacity and $\tau_z$ is the optical depth such that
\begin{equation}
    \tau_z = \int_z^{\infty} \kappa \rho \mathrm{d}z'.
\end{equation}
The total luminosity could then be computed over the two disk surfaces as
\begin{equation}
    \label{eq:lum}
    L = 2\iint l \mathrm{d}A\,.
\end{equation}
However, without live-cooling, we find that the regions that will dominate the emission are layers of hot, low-density gas in the upper atmosphere, which should also be actively cooling. Figure~\ref{fig:tau} depicts the gas temperature and optical depths for the fiducial simulation at late times. We see that there is an artificially hot layer at large $z$ that will dominate the emission, even though it should have rapidly cooled due to its very low optical depth. 

To provide a rough estimate of the luminosity $L_\mathrm{est}$, we can use Equation~\ref{eq:lum}, but limit the integration for $l$ to $\tau > 10$, so avoiding the problematic regions in the upper disc atmosphere which are not faithfully modelled in our simulation. We model the absorption opacity considering contributions from Kramers, $H^-$ opacity and electron scattering (see Appendix~\ref{sec:opacity}). Under this treatment we calculate luminosities of lone BHs and of BHs undergoing close encounters of $L_\mathrm{est} \sim 10^{-3} L_\mathrm{AGN}$ and $L_\mathrm{est} \sim 10^{-4}L_\mathrm{AGN}$ respectively. This is super-Eddington compared to the mass of the embedded objects, equivalent to approximately $6L_\mathrm{Edd,BH}$ and $70L_\mathrm{Edd,BH}$ respectively. Super-Eddington emission is not necessarily unphysical as the heating is driven by shocks and the system is not in steady state. We note that these estimates are very sensitive to the threshold imposed on the minimum optical depth; if integration is limited to $\tau < 5$, the luminosities rise by a factor 100. Even if cooling could be accounted for, the true system luminosities remain highly uncertain, as if radiation pressure is included, we might observe reduced gas temperatures at equivalent pressures which could lead to a reduction in the observed luminosity. While precise calculations of the novae luminosities are beyond our reach, the existence of minidisc collisions at periapsis and the associated injection of thermal energy into the Hill sphere suggest the total luminosity should rise periodically with each periapsis passage. Without knowledge of the cooling times for the novae shells when they reach the upper atmosphere, it is hard to predict how the luminosity should fall after each peak, complicating the emission profile once the binary hardens and the frequency of novae generation increases. The results of our work here do not rule out potential electro-magnetic signatures associated with binary formation in AGN discs. We leave a more precise analysis of the thermal emission to future studies, ideally those including live radiative transfer in the simulation.

\section{Discussion}
\label{sec:discuss}

\subsection{Consequences for the AGN Channel}
\label{sec:consq}

AGN offer an incredible range of different environments for binary interactions to take place in, as such it is hard to make sweeping statements about the general likelihood of binary formation without better knowledge of where exactly in each AGN disc these interactions take place, or even in what type of AGN disc they are likely to be embedded in. That said, our results indicate that binary formation is still viable for a variety of environments, and should be very likely in the outer regions of the AGN disc where the large Hill spheres allow for significant mass accumulation about each BH. Recently, \citep{Rowan_2025a} used Monte-Carlo simulations to explore how the presence of gas would effect the merger rates of BHs embedded in a variety of different AGN. They found the bottleneck to be the alignment time of BHs: the time taken for the AGN disc to embed initially inclined BHs within the disc midplane. This alignment timescale is generally shorter for BHs more distant from the SMBH. Taking our own result and the results of \citet{Rowan_2025a} in hand, we might predict an increase in overall binary formation efficiency in AGN due to BHs being more likely to interact in the outer AGN disc, where formation is easier. Regardless, with the most realistic numerical treatment to date, we are able to reaffirm the ability of BHs to form binaries in AGN discs without the need for extreme AGN disc environments. We are unable to reliably comment on the likelihood of binary formation in the far outer reaches of the AGN discs, where star formation is required to maintain gravitational stability. Simulations of these environments will require a better understanding of the initial conditions of such systems, and potentially the inclusion of more advanced physical processes such as gas self-gravity.

\subsection{Magnetic Cyclones}
\label{sec:mag_cyc}

\begin{figure}
    \includegraphics[width=\columnwidth]{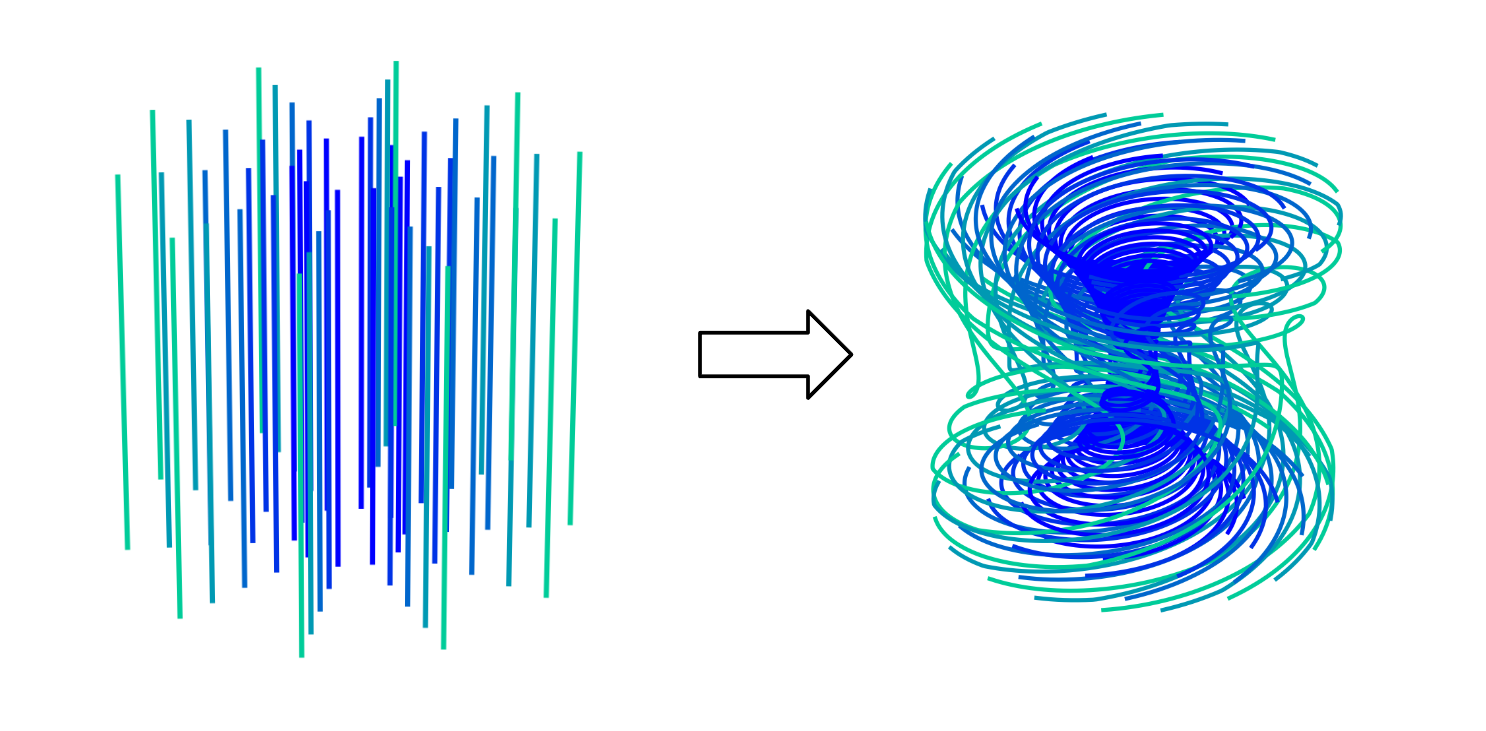}
    \caption{Exaggerated schematic description for how a combination of vertical differential rotation and advection may result in flux rope formation from a magnetic field initially aligned with the $\hat{\bm{z}}$ axis. In real systems, the flow velocity will be much more complicated, and magnetic pressure may have a significant back-reaction on the winds. Despite this, some degree of field winding should be expected from the counter-rotating cyclonic winds.}
    \label{fig:field}
\end{figure}

Magnetic fields have been neglected from this study, primarily due to the considerable increase in computational complexity associated with running magneto-hydrodynamic simulations. However, with the introduction of counter-rotating cyclonic winds, consideration as to the potential effects on the local magnetic field structure is warranted. Specifically, the differential rotation has the potential to twist up any pre-existing magnetic field structure resulting in the formation of a flux rope. When combined with the inwards advection of field lines in the minidisc and the outwards advection in the wind, we might expect significant magnetic field deformation and concentration near the BHs. Figure~\ref{fig:field} represents a schematic description of how this rotation and advection might deform a pre-existing vertical magnetic field. Such deformations may enhance the strength of the wind by introducing a vertical magnetic pressure gradient, and may even lead to the formation of jet-like structures. If the field is well-entrained with the flow, the coalescence of the minidisc winds as the BHs approach close encounter will crush anti-aligned flux ropes against each other, allowing for the potential reconnection. These reconnection events may produce significant electromagnetic emission, giving a new signature associated with binary formation.

It is clear that the prediction of counter-rotating winds presents a new environment for novel magneto-hydrodynamic phenomena. However, without live simulation with magnetic fields, we are unable to account for the feedback of magnetic forces on the gas or any inefficiencies in advection of the field by the gas. Such work goes beyond the scope of this study, but we are hopeful that future studies into this area may uncover new observable signatures of embedded black holes in AGN either in isolation or during binary formation.

\subsection{Comparison to Literature}

This study is a continuation to the recent developments into the simulation of compact objects embedded in AGN disc. It marks the first time that a non-isothermal equation--of-state has been used to simulate embedded BHs in 3D. We compare the results of this paper to similar studies in the field. The most direct comparison can be made to \citetalias{Whitehead_2024II} which considered similar systems in 2D\footnote{This previous study also included contributions from radiation, which we neglect here.}. Our findings generally support the results of this earlier paper, showing that the same disc novae are generated. When modelled in 3D, the blasts preferentially escape vertically away from the midplane, resulting in a less severe disruption of the minidiscs. We also document thermal winds from the minidiscs in isolation which were not able to be modelled in 2D. As noted in \citetalias{Whitehead_2024II}, the hydrodynamics of our system bear some resemblance to the embedded supernovae simulations of \citet{Grishin_2021} due to the strong vertical outflows present in both. The outbreak of the supernova shock from the disc shows morphological similarities to the wind outbreak in our paper, but we note that the supernova shock propagates significantly further and on a much shorter timescale. Other studies have considered analytical models for electromagnetic emission associated with compact objects in AGN, either post-merger \citep{McKernan_2019, Graham_2020, Kimura_2021, Wang_2021, Tagawa_2023_merge}
or for a single BH in isolation \citep{Tagawa_2023_single}. We hope the confirmation of novel wind and novae phenomena associated with embedded compact objects helps motivate further research into this area. 

We have shown that rather than a disc structure, the gas within the Hill sphere for embedded isolated BHs is approximately spherically symmetric. The density and temperature radial profiles are convective unstable, resulting in a system akin to a star with a very compact BH core and a low mass convective envelope. Many recent studies have considered the formation and evolution of massive stars within the discs of AGN \citep{Goodman_2004, Cantiello_2021, Jermyn_2021, Dittmann_2021, Jermyn_2022, Chen_2024a, Fabj_2024, Chen_2024b}. Such objects are of specific interest to the AGN channel for gravitational wave formation, as they can leave behind compact object remnants with high mass and spin \citep{Jermyn_2021}. Understanding exactly how the BH stars documented in this study compare to these AGN stars will require more advanced modelling of the ``stellar'' evolution, such as accretion onto, and feedback from, the central BH, nuclear fusion, radiative processes and mass loss by wind.

When considering the likelihood on binary formation, we find that our previous isothermal 2D study \citepalias{Whitehead_2024I} overestimates the strength of gas dissipation. Direct comparison is harder to make as the gas densities explored in that 2D study were substantially lower. However, if the 2D isothermal dissipation model (see Equation 24 of \citetalias{Whitehead_2024I}) is extrapolated to the systems explored in this paper, we find the older model predicts dissipations around an order of magnitude higher than documented here. Similar issues were noted in \citetalias{Whitehead_2024I} and \citetalias{Whitehead_2024II} which found both a weakening of the fit and overestimation of dissipation at higher densities. Other hydrodynamical studies have offered models for binary formation likelihood, considering the AGN disc mass scale and energy before close encounter \citep{Li_2023} or the Hill impact parameter and energy before close encounter \citep{Rowan_2024}. Given the wide variety of AGN environments and BH trajectories studied in this paper, along with the more advanced hydrodynamical treatment, we expect the formation criteria proposed in Section~\ref{sec:model} to represent the most accurate modelling available for binary black hole formation in AGN. 

\subsection{Caveats and Limitations}

In this study we have adopted various assumptions that should be carefully considered when interpreting the results.

\subsubsection{Gas Physics}
This study marks the first attempt to study gas assisted binary BH formation in AGN using both a 3D hydrodynamic domain and a non-isothermal equation-of-state. However, while the treatment adopted here is a marked improvement over an isothermal prescription, there are many physical processes that effect the hydrodynamical evolution that we have not included. Foremost among these are:
\begin{itemize}
    \item \textit{Radiation and cooling}: we have neglected the effects of radiation pressure in this study. Avoiding solving the radiation evolution equations reduces the computational cost significantly, but results in less accurate simulation where density is low and temperature is high, such as far above the midplane. Most importantly, without radiation transport we are unable to include live cooling in our simulations and cannot make accurate predictions as to the luminosity of the system. 
     \item \textit{Viscosity, viscous heating, and viscous feedback}: following \citetalias{Whitehead_2024II} we set the viscosity to its initial value in the ambient gas-pressure dominated disk and neglect variability in space and time. Higher viscosities could lead to more aggressive heating close to the BHs. 
    \item \textit{Magnetism}: for reasons of computational cost, we have neglected the effects of magnetic fields from this study. The introduction of a counter-rotating wind to this system clearly motivates future consideration as to the potential magnetic field structure (see Figure~\ref{fig:field}), but it is not yet clear how the introduction of magnetic feedback on the gas will affect the minidisc or wind structure.
    \item \textit{Gas Self-Gravity}: in the outer AGN disc gas masses in the Hill sphere during close encounter can become comparable to the mass in the BH masses, potentially resulting in significant gas self-gravity forces. 
\end{itemize}

\subsubsection{Limited Parameter Space}

The significant increase in cost associated with moving from 2D to 3D prevents us from making as wide a parameter study as we might like. We have been forced to select only 9 different disc environments in which to study formation events and while each environment was ran with 15 impact parameters, the fractal nature of the hierarchical 3-body system means that a very fine spacing in $b$ is required to study all encounter types. Moreover, these trajectories were limited to initially circular orbits in the midplane of the AGN disc, meaning no investigation could be made as to the effects of pre-encounter inclination, eccentricity, or velocity. As such we are unable to draw quantitative conclusions as to the efficiency of binary BH formation for more general interactions in AGN discs. Caution is warranted when propagating the conclusions of this study to a wider population of AGN environments and encounter types. 

\subsubsection{BH Physics}
Our simulations are scaled with respect to the Hill sphere and so vary with $R_0$, but even for our smallest systems with $R_0=5\times 10^3 R_g$, the minimum cell size and smoothing length are approximately $2.5\times 10^4 r_g$ and $9.4\times 10^5 r_g$ respectively (here $r_g$ is the Schwarzchild radius for a $25M_\odot$ BH). As such, the physical accretion surface of the BH always lies many orders of magnitude beneath the length scales resolvable by our simulation. Hence there exists a sub-grid BH minidisc whose effects are not accounted for in our system. We also do not include any accretion or BH feedback effects such as jets, which could change the gas morphology within the Hill sphere. As discussed in Section~\ref{sec:bh_star}, the total thermal energy within the Hill sphere is dependent on the softening length or, in real systems, the inner radius at which the power-law energy profile breaks. While Eddington-limited accretion is unlikely to have a significant effect on the total mass within the Hill sphere, energetic feedback could be thermally significant depending on the inner radius at which the power-law profiles no longer hold. The decision to neglect the effects of accretion is tentatively supported by the results of \cite{Rowan_2024} which found that the energy dissipated in accretion-less isothermal BH-BH encounters was in agreement within a factor of 2-3 compared to those systems with accreting BHs (when the softening length was equal to the sink radius).. The accuracy of sub-grid accretion prescriptions remain an open question in this field and care is warranted when considering the physical motivation for such routines. Further work is required in this field to determine fully the effects of accretion on binary formation.

\section{Summary and Conclusions}
\label{sec:conclusions}

In this work we have simulated 135 binary interactions between initially isolated embedded BHs in 3D, using an adiabatic hydrodynamical treatment. We have analysed the circum-BH gas structure in isolation and during close encounters, affirming the existence of previously predicted disc novae and showing for the first time the thermal wind structure driven by the embedded minidisc. We summarise the key findings below:

\begin{itemize}
    \item Shock heating by infalling gas leads to the generation of strong thermal winds that emanate upwards and outwards from the minidisc. This results in a minidisc half as massive as observed in 2D studies (Figure~\ref{fig:wind_growth}).
    \item The distribution of gas within the Hill sphere is approximately spherically symmetric, comparable to a star with a BH core and a low mass convective envelope. The gas properties are well modelled by power laws in radius (Equations~\ref{eq:ss_rho}-\ref{eq:ss_P}).
    \item The minidisc winds rotate in the opposite direction to the minidiscs, with interesting consequences for the local magnetic field structure potentially resulting in the formation of flux ropes and driving reconnection during binary interactions (Figure~\ref{fig:field}).
    \item We confirm the existence of disc novae in 3D, with large thermal blasts generated by the collision of minidiscs during BH close encounters (Figure~\ref{fig:static_collision}).
    \item We show that binary formation from single-single encounters is viable in 3D with a non-isothermal equation of state. Binary formation is most likely in regions of high Hill mass, generally found in the outer regions of the AGN due to the increased size of the Hill sphere (Figure~\ref{fig:outcomes}). 
    \item We show that for our initial conditions, to achieve a binary formation likelihood of $(5\%, 20\%, 50\%, 100\%)$, ambient Hill masses of $m_\mathrm{H,0}\sim(0.01, 0.025, 0.063, 0.2)\mbh$ are required (Figures~\ref{fig:mass} and~\ref{fig:enrichment}).
    \item We provide a comprehensive overview of the different factors affecting the likelihood of binary formation, considering the effect of different trajectories and the involvement of the SMBH in each of these complex triple interactions (Figure~\ref{fig:arguments}). 
    \item We provide fitting formulae for the binary energy dissipated by gas during close encounters, and suggest a statistical methodology for predicting the likelihood of binary formation based on the ambient Hill mass $m_{\mathrm{H},0}$ and depth of first periapsis $\rp$ (Equations~\ref{eq:model_A} and~\ref{eq:model_B}).
\end{itemize}

In simulating a comprehensive suite of binary black hole encounters in a variety of disc environments under the most realistic hydrodynamic treatment to date, we continue to advance our modelling and understanding of binary formation in AGN.

\section*{Acknowledgements}
The simulations presented in this paper were performed using resources provided by the Cambridge Service for Data Driven Discovery (CSD3) operated by the University of Cambridge Research Computing Service \href{https://www.csd3.cam.ac.uk}{(www.csd3.cam.ac.uk)}. Funding for CSD3 usage was provided by UKRI through opportunity OPP503 as application APP35272. Preparatory simulations were performed on the \texttt{Hydra} cluster at The University of Oxford. This work was supported by the Science and Technology Facilities Council Grant Number ST/W000903/1.

\section*{Data Availability}

The data underlying this article will be shared on reasonable request
to the corresponding author.



\bibliographystyle{mnras}
\bibliography{citations} 




\appendix
\label{sec:appendices}

\section{Analytical Power-laws}
\label{sec:virial}
To derive the analytic power-law radial profiles given in Equations~\ref{eq:ss_rho}-\ref{eq:ss_P}, we consider a spherically symmetric hydrostatic distribution of gas about an isolated BH with mass $\mbh$, neglecting gas self-gravity and any frame forces. Satisfying hydrostatic equilibrium requires
\begin{align}
    \label{eq:app_hstatic}
    \partial_r P + \rho(r)\partial_r \phi(r) = \partial_r P + \frac{G\mbh\rho(r)}{r^2} &= 0
\end{align}
If we model the gas as adiabatic and ideal such that $(\gamma-1)P = U$ where $U$ is the internal energy, we can multiple Equation~\ref{eq:app_hstatic} by $r^3\mathrm{d}r$ and integrate over the Hill sphere to derive the virial theorem
\begin{equation}
    \label{eq:virial}
    3(\gamma - 1)K_\mathrm{H} + V_\mathrm{H} = 0
\end{equation}
where $K_\mathrm{H}$ and $V_\mathrm{H}$ are the total thermal and potential energy within the Hill sphere defined as
\begin{equation}
    \label{eq:app_thermal}
    K_\mathrm{H} = \int_0^{\rhs} U(r) \cdot 4\pi r^2 \mathrm{d}r 
\end{equation}
\begin{equation}
    \label{eq:app_potential}
    V_\mathrm{H} = \int_0^{\rhs} \phi(r)\rho(r) 4\pi r^2 \mathrm{d}r 
\end{equation}
We now seek power law solutions to Equation~\ref{eq:app_hstatic}, such that $\rho(r) = \tilde{\rho}r^a$ and $T(r) = \tilde{T}r^b$. It is straightforward to show that hydrostatic equilibrium requires
\begin{align}
    b &= -1 \\
    \tilde{T} &= \frac{G\mbh \mu m_p}{(1-a)k}
\end{align}
To constrain $a$ and $\tilde{\rho}$ we substitute our power law profiles for $\rho$ and $T$ (and therefore also $P$ and $U$) into the expressions for the thermal and potential energy (Equations~\ref{eq:app_thermal} and~\ref{eq:app_potential}) to find
\begin{equation}
    \label{eq:app_thermal2}
    K_\mathrm{H} = \frac{4\pi G\mbh \tilde{\rho}}{(1-a)(\gamma-1)}\int_0^{\rhs} r^{a+1}\mathrm{d}r
\end{equation}
\begin{equation}
    \label{eq:app_potential2}  
    V_\mathrm{H} = - 4\pi G \mbh \tilde{\rho}\int_0^{\rhs} r^{a+1}\mathrm{d}r
\end{equation}
For these forms to satisfy virial equilibrium (Equation~\ref{eq:virial}), we require $a=-2$. We can convert $\tilde{\rho}$ into a more useful form using the definition of the Hill mass $m_\mathrm{H,s}$.
\begin{equation}
    m_\mathrm{H,s} \equiv \int_0^{\rhs} \rho(r) \cdot 4\pi r^2 \mathrm{d}r = 4\pi \tilde{\rho} \rhs
\end{equation}
The analytic forms for the gas radial profiles are thus
\begin{align}
    \rho(r) &= \frac{m_\mathrm{H,s}}{4\pi \rhs}r^{-2} \\
    T(r) &= \frac{G\mbh \mu m_p}{3k}r^{-1} \\
    P(r) &= \frac{G\mbh m_\mathrm{H,s}}{12\pi \rhs}r^{-3} \\
    U(r) &= \frac{G\mbh m_\mathrm{H,s}}{8\pi \rhs}r^{-3}
\end{align}
While it is not obvious that the assumptions made here (hydrostatic, virialised, spherically symmetric) should apply to the full hydrodynamic simulation, the analytic forms match the simulation profiles exactly. It is worth noting that with $a=-2$, the energy integrals in Equations~\ref{eq:app_thermal2} and~\ref{eq:app_potential2} diverge; this issue does not manifest in the hydrodynamic simulations as the gravitational potential is smoothed for $r<h=0.025\rhs$, resulting in a break from the power-law profiles (see Figure~\ref{fig:radial}). Even for an unsmoothed gravitational potential, the finite resolution of the simulation would also prevent this divergence.

\section{Locations of Jacobi Regions}
\label{sec:jac_arg}
\begin{figure*}
    \includegraphics[width=2\columnwidth]{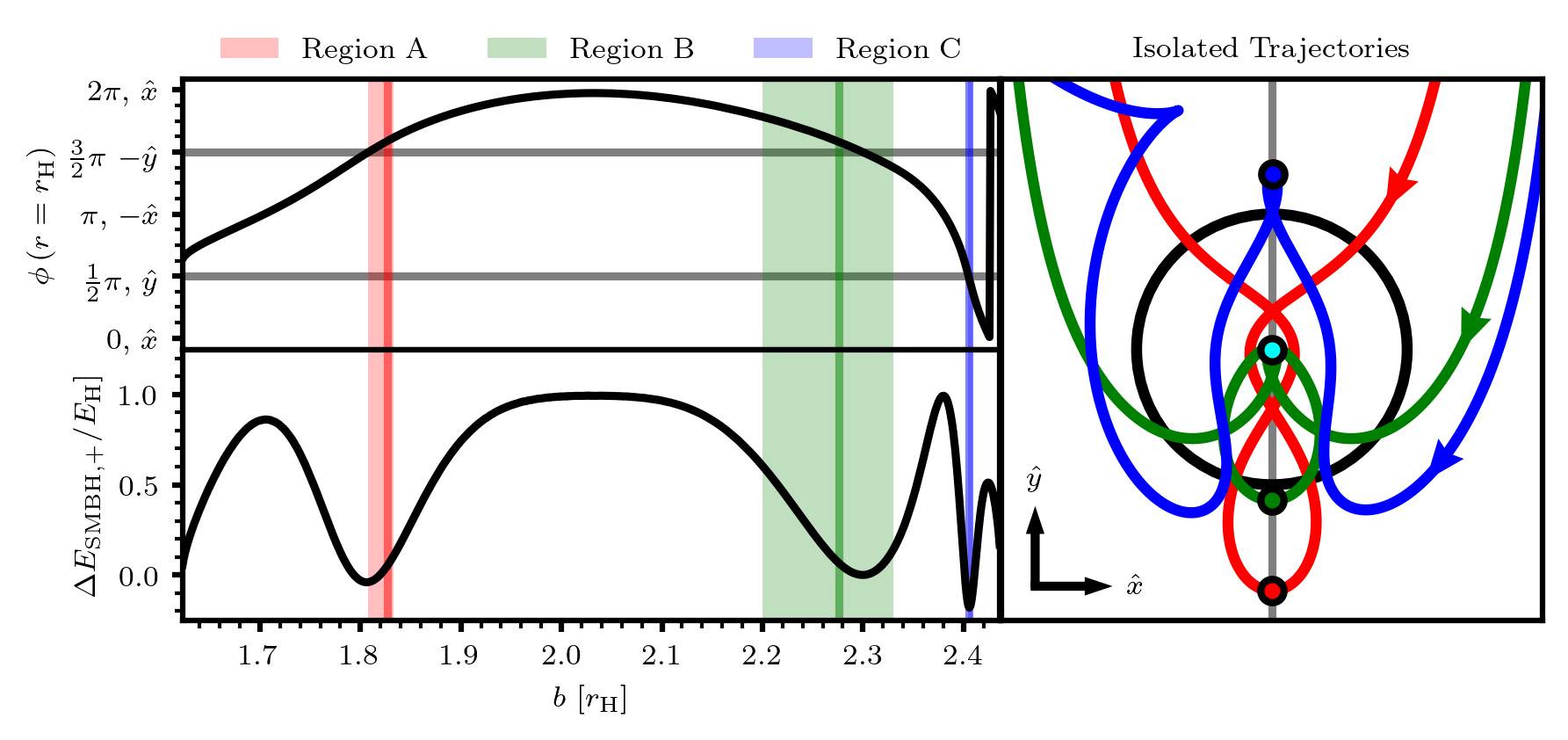}
    \caption{Variation in the argument of first exit from the Hill sphere $\phi\left(r=\rh\right)$, and the energy injected by the SMBH post-periapsis $\Delta E_{\mathrm{SMBH},+} = \int_{r_p}^{\rh}\epsilon_\mathrm{SMBH} \mathrm{d}t$ as a function of impact parameter for 1000 gas-less flybys. The argument is the angle made between the binary separation vector and the $\hat{\bm{x}}$ axis. The right panel shows the three trajectories with apoapsides most-aligned with the $y$ axis. These trajectories are highlighted as vertical lines in the left panels. On these most-aligned trajectories, the SMBH does effectively no work; the minima in $\Delta E_{\mathrm{SMBH},+}$ are found where $\phi(r=\rh) \sim \frac{\pi}{2}$ or $\frac{3\pi}{2}$. Argument of first exit is used instead of argument of apoapsis, as not all trajectories feature an apoapsis. This results in a shift in $b$ between the minima of the SMBH work calculated for the left panel and the true maximally aligned trajectories depicted in the right panel.}
    \label{fig:ngas_arg}
\end{figure*}

It has been previously shown that in the absence of gas, hierarchical triples can exhibit unusual behaviour where the most massive tertiary component forces the inner binary to perform multiple interactions before separation \citep{Goldreich+2002,Boekholt_2022}. These interactions are called Jacobi encounters, and occur for specific ranges of impact parameters $b$ (see Figure~\ref{fig:jacobi}). What was not previously obvious is that trajectories in all three Jacobi regions share a common factor; post-periapsis they all feature trajectories closely aligned with the the positive and negative $\hat{\bm{y}}$ axis. This is because on these trajectories, the ionising work done by the SMBH is reduced, as the centrifugal action acts along the $\hat{\bm{x}}$ axis. If the apoapsis of the first trajectory is closely aligned with the $\hat{\bm{y}}$ axis, the SMBH does effectively no work during the period and so the system can perform multiple encounters. To illustrate this, we include Figure~\ref{fig:ngas_arg} which depicts the 3 most-$y$-aligned trajectories and their membership within each of the Jacobi regions. 

\section{Opacity Modelling}
\label{sec:opacity}
We detail briefly the methodology used to approximate the opacity of the gas when estimations as to the system luminosity are made. The opacity due to free-free, bound-free and bound-bound electron transitions are modelled using Kramers formula
\begin{equation}
    \kappa_K \simeq 4\times 10^{25} \left(1+X\right)\left(Z+0.001\right)\rho T^{-\frac{7}{2}} \;\mathrm{cm}^2\mathrm{g}^{-1},
\end{equation}
where we assume $Z \sim 0.01$ and maintain $X=0.7$ as the metal and hydrogen mass fractions respectively. We also consider opacity due to the negative hydrogen ion $H^-$
\begin{equation}
    \kappa_{H^-} \simeq 1.1\times 10^{-25} Z^\frac{1}{2}\rho^\frac{1}{2}T^{7.7} \mathrm{cm}^2\;\mathrm{g}^{-1}.
\end{equation}
There is also a flat contribution from electron scattering
\begin{equation}
    \kappa_\mathrm{es} = 0.2\left(1+X\right)\;\mathrm{cm}^2\mathrm{g}^{-1}.
\end{equation}
We can crudely approximate the combinations of these opacities over a range of temperatures using
\begin{equation}
    \kappa \simeq \frac{1}{\frac{1}{\kappa_{H^-}} + \frac{1}{\kappa_\mathrm{es} + \kappa_K}}
\end{equation}
This is the opacity used to approximate the luminosities in Section~\ref{sec:observables}.


\bsp	
\label{lastpage}

\end{document}